\documentclass[fleqn,usenatbib]{mnras}

\usepackage[T1]{fontenc}
\usepackage{ae,aecompl}

\usepackage{graphicx}	
\usepackage{amsmath}	
\usepackage{amssymb}	
\usepackage{multicol}
\usepackage{booktabs,tabularx}
\usepackage{enumitem}
\usepackage{multirow}

\newcommand{\ee}[1]{\times 10^{#1} } 
\newcommand{\hst}{\textit{HST}} 
\newcommand{\swift}{\textit{Swift}} 
 
\newcommand{\msun}{\text{M}_\odot}

\title[Looking under the lamppost]{Using AGN lightcurves to map accretion disc temperature fluctuations}

\author[Neustadt \& Kochanek]{
J.~M.~M.~Neustadt$^{1}$\thanks{E-mail: neustadt.7@osu.edu (JMMN)} and
C.~S.~Kochanek$^{1,2}$ 
\\
$^{1}$Department of Astronomy, The Ohio State University, 140 West 18th Avenue, Columbus, OH 43210, USA \\
$^{2}$Center for Cosmology and AstroParticle Physics (CCAPP), The Ohio State University, 191 W. Woodruff Avenue, Columbus, OH 43210, USA \\ 
}

\date{Accepted XXX. Received YYY; in original form ZZZ}

\pubyear{2022}

\begin{document}
\label{firstpage}
\pagerange{\pageref{firstpage}--\pageref{lastpage}}
\maketitle

\begin{abstract}

We introduce a new model for understanding AGN continuum variability. We start from a Shakura--Sunyaev thin accretion disc with a steady-state radial temperature profile $T(R)$ and assume that the variable flux is due to axisymmetric temperature perturbations $\delta T(R,t)$. After linearizing the equations, we fit UV-optical AGN lightcurves to determine $\delta T(R,t)$ for a sample of seven AGNs.  We see a diversity of $|\delta T/T| \sim 0.1$ fluctuation patterns which are not dominated by outgoing waves traveling at the speed of light as expected for the ``lamppost'' model used to interpret disc reverberation mapping studies.  Rather, the most common pattern resembles slow ($v \ll c$) ingoing waves.  An explanation for our findings is that these ingoing waves trigger central temperature fluctuations that act as a lamppost, producing lower amplitude temperature fluctuations moving outwards at the speed of light. The lightcurves are dominated by the lamppost signal -- even though the temperature fluctuations are dominated by other structures with similar variability time-scales -- because the discs exponentially smooth the contributions from the slower moving ($v \ll c$) fluctuations to the observed lightcurves. This leads to lightcurves that closely resemble the expectations for a lamppost model but with the slow variability time-scales of the ingoing waves. This also implies that longer time-scale variability signals will increasingly diverge from lamppost models because the smoothing of slower-moving waves steadily decreases as their period or spatial wavelength increases. 

\end{abstract}

\begin{keywords}
accretion, accretion discs -- galaxies: active -- methods: data analysis 
\end{keywords}

\section{Introduction}

The stochastic continuum variability of active galactic nuclei (AGNs) has been studied for decades (e.g., \citealt{oknyanskij78,perola82,ulrich97,cristiani97,giveon99,geha03}).  Quantitatively, they can be reasonably well-described as modest-amplitude damped random walks \citep{kelly09,kozlowski10,macleod10,macleod12,zu13}, although this description is not exact (e.g., \citealt{kelly11,mushotzky11,zu13}).

Fundamentally, the flux variability should be driven by temperature fluctuations in the accretion disc surrounding the supermassive black hole (SMBH).  Because shorter wavelengths are generally observed to vary first with lags between wavelengths typical of the light travel time across a disc (e.g., \citealt{sergeev05,cackett07}), the ``lamppost model'' is in common use.  In this model, fluctuations in the luminosity of the central region illuminate the outer regions and drive temperature fluctuations in the disc which in turn drive the variability.  This assumption is often used for ``disc reverberation mapping'' where the interband lags are used to constrain the temperature profile of the disc (e.g., \citealt{shappee14,fausnaugh16,edelson17,vincentelli21}), similar to broad line region (BLR) reverberation mapping (RM) between the continuum and the emission lines on longer time-scales \citep{blandford82,peterson93}.  

Frequently, the central source is ascribed to X-ray emission \citep{nayakshin00,frank02}.  However, there are cases where the X-rays vary after the UV/optical or show uncorrelated structures that call this assumption into question (e.g., \citealt{berkley00,kazanas01,edelson19,dexter19}).  In most studies of disc RM, the model is generally only invoked to justify the lag interpretation rather than as a detailed physical model.  

There have been studies that argue against the lamppost model.  For example, \citet{dexter11} argue that disc variability is largely due to inhomogeneous temperature fluctuations.  Others argue for intrinsic thermal fluctuations in the disc as the origin of the variability, based on the fact that the observed variability time-scales are typical of the thermal time-scales at the disc radii producing the observed flux (e.g., \citealt{kelly09,burke21}).  These long time-scales are a significant problem for the lamppost model, as they are much longer than any characteristic time-scale associated with the very inner regions of the disc.  

Here we attempt to resolve this discrepancy between the expected disc-driven variability and the observed lamppost-like signals.  We develop a model that tracks the temperature fluctuations over time as a function of disc radius by modeling AGN lightcurves obtained over a broad range of wavelengths.  Because the temperature of the disc changes with radius, the different filters of the lightcurves correspond to different radii. In the thin disc model, temperature decreases with increasing radii, and so decreasing the effective wavelength of a filter corresponds to probing smaller radii.

This model can serve as a test of the lamppost model.  If the lamppost model is the source of all flux variations in the disc, it follows that any temperature fluctuations should move radially outward at the speed of light.  Slower or inward moving waves would be evidence for other coherent sources of variability, and incoherent patterns would be evidence for a model like \citet{dexter11}.  A limitation of our model is that we must assume axisymmetry to make it mathematically tractable.  Tightly wound spiral waves will be well approximated by axisymmetry, but the completely stochastic and inhomogeneous model of \citet{dexter11} will not be.  We must also include some smoothing parameters, and we will explore their effects in detail. 

In Section~\ref{sec:methods}, we discuss how we build the model, and in Section~\ref{sec:testing}, we show how the model works with a range of test problems.  In Section~\ref{sec:ngc5548}, we apply the model to NGC~5548 considering a broad range of parameter choices, and in Section~\ref{sec:others}, we apply the model to other AGNs using a more limited range of model parameters.  In Section~\ref{sec:discussion}, we discuss our conclusions, interpretations, and consider how a lamppost signal can dominate the lightcurves while not being the dominant source of temperature fluctuations.


\section{Methods}\label{sec:methods}

Here we introduce the main equations that govern our model, going from the steady-state radial temperature profile $T(R)$ to temperature fluctuations over time and radius $\delta T(R,t)$, including additional factors due to light travel time.   Next, we will discuss the linear inversion process that allows us to construct $\delta T(R,t)$ from the lightcurves. We will then discuss how the physical parameters of the AGNs change the range of disc radii probed by our model.  Finally, we discuss the AGN lightcurves that we will examine with our model.  We do not include any relativistic effects (beaming, gravitational redshifts, Doppler shifts, raybending, etc.) as these effects are generally weak for the radii producing the observed data.

\subsection{Equations}

The accretion discs of AGNs are often modeled using the Shakura--Sunyaev thin disc model \citep{shakura73}, with a radial temperature dependence of $T \propto R^{-3/4}$, yielding a UV/optical SED with flux $F_\nu \propto \nu^{1/3}$ (alternatively, $F_\lambda \propto \lambda^{-7/3}$). Using a different steady-state temperature profile would only shift the location and scale of temperature fluctuations rather than change their overall structure.  For the unperturbed disc we use this thin disc model with an inner radius 
\begin{equation}
R_\text{in} =  \alpha R_g = \frac{\alpha G M_\text{BH}}{c^2} 
\end{equation}
where $R_g$ is the gravitational radius. We will assume $\alpha=6$, the innermost stable orbit of a Schwarzchild BH.  The steady-state temperature profile of a thin disc as function of the dimensionless radial variable $u = R/R_{\rm in}$ is 
\begin{equation}
T_0(u) = T_\text{in}  u^{-3/4} (1-u^{-1/2})^{1/4}
\end{equation}
where
\begin{equation} \label{eq:temp_in}
\begin{split}
T_\text{in} & = \bigg(\frac{3 G M_\text{BH} \dot{M}  }{8 \upi \sigma R_\text{in}^3}\bigg)^{1/4} \\
& = 1.54\ee{5}~\text{K} ~ \bigg(\frac{L}{L_\text{Edd}}\bigg)^{1/4}
\bigg(\frac{10^9~M_\odot}{M_\text{BH}}\bigg)^{1/4}  \bigg(\frac{6}{\alpha}\bigg)^{3/4}
\end{split}
\end{equation}
and 
\begin{equation}
\dot{M} = \frac{L_{\rm Edd}}{\eta c^2}  \bigg(\frac{L}{L_{\rm Edd}}\bigg) 
\end{equation} 
where $L_{\rm Edd}$ is the Eddington luminosity.  We will assume an accretion efficiency $\eta = 0.1$ and define $\lambda_{\rm Edd} = L/L_{\rm Edd}$. 

The flux of the accretion disc at wavelength $\lambda$ is 
\begin{equation}
F_\lambda = \frac{2\upi R_{\rm in}^2 \cos{i} }{D^2} \int_1^\infty ~ u du ~ B_\lambda(T_0(u)) 
\end{equation}
where $D$ is the distance, $i$ is in the inclination of the disc to our line of sight, and $B_\lambda$ is the Planck function.  This can be rewritten as
\begin{equation}
F_\lambda = F_{\lambda,0}(\lambda) \int_1^\infty \frac{u du}{e^x-1}  
\label{eq:flux}
\end{equation}
where
\begin{equation}
F_{\lambda,0}(\lambda) = \frac{4 \upi h  c^2 \cos{i} R_\text{in}^2}{\lambda^5 D^2} 
\end{equation}
and
\begin{equation}
 x = \frac{hc}{\lambda k_B T_0(u)} ~.
\end{equation}
Assuming linear perturbations in the temperature $\delta T (u,t)$, the change in flux is 
\begin{equation} \label{eq:dfdt}
\delta F(t) = F_{\lambda,0}(\lambda) \int \frac{u du}{(e^x-1)^2}  ~\frac{x e^x}{T_0(u)} ~ \delta T (u,t) ~.
\end{equation} 
If we grid the disc in radius $u_k$ and time $t_l$ then the change in flux for wavelengths $\lambda_j$ can be discretized into matrix $W$ as 
\begin{equation} \label{eq:linear}
\delta F(\lambda_j,t_l) = W(\lambda_j,u_k) \delta T(u_k,t_l) 
\end{equation}
where the matrix elements are evaluated as radial integrals across each bin.

Because of the size and inclination of the accretion disc, we must consider time delays due to light travel time between different parts of the disc.  Flux changes emitted from any radius at the same time in the SMBH reference frame will be smeared over a range of observed times.  The effect of this is that flux changes at radius $u$ emitted at model parameter time $t_p$ have weighted contributions to a range of observed data times $t_d$ ($t_p$ for \textit{p}arameter time, $t_d$ for \textit{d}ata time) of the form $f(u,t_p,t_d)$.  This modifies Equation~\ref{eq:linear} to
\begin{equation}
\delta F(\lambda_j,t_d) = \Big[ W(\lambda_j,u_k) \cdot f(u_k,t_p,t_d) \Big] \delta T(u_k,t_p) ~.
\end{equation}
The characteristic time-scale of the smearing is the light travel time corrected for inclination
\begin{equation} \label{eq:t0}
 t_0 = \frac{R}{c} \sin{i} = \frac{u R_{\rm in}}{c} \sin{i} ~.
\end{equation} 
If we define the functions
\begin{equation}
\begin{split}
 G_1(t_\alpha,t_\beta) = & ~\frac{1}{\upi \Delta t} \sqrt{(t_0^2 - t_\beta^2)^2 - (t_0^2 - t_\alpha^2)^2} ~,~ {\rm and } \\
 G_2(t_\alpha,t_\beta) = & ~\frac{1}{\upi} \bigg[ \arcsin{\bigg(\frac{t_\beta}{t_0}\bigg)} - \arcsin{\bigg(\frac{t_\alpha}{t_0}\bigg)} \bigg] \\
 \end{split}
\end{equation}
then
\begin{equation}
\begin{split}
 f(u,t_p,t_d) = &~ G_1(t_1,t_2) + \frac{t_d-(t_p - \Delta t)}{\Delta t} G_2(t_1,t_2) \\
 + & G_1(t_3,t_4) + \frac{(t_p+\Delta t)-t_d}{\Delta t} G_2(t_3,t_4)
\end{split}
\end{equation}
where 
\begin{equation}
\begin{split}
t_1 & ~= \max{\{t_p-\Delta t-t_d,-t_0\}} ~, \\ 
t_2 & ~= \min{\{t_p-t_d,+t_0\}} ~, \\ 
t_3 & ~= \max{\{t_p-t_d,-t_0\}} ~, \\ 
t_4 & ~= \min{\{t_p+\Delta t-t_d,+t_0\}} ~,
\end{split}
\end{equation}
and $\Delta t$ is the grid spacing in time.  If we absorb $f$ into $W$, we can write out our final expression as:
\begin{equation} \label{eq:final_matrix}
\delta F(\lambda_j,t_d) = W(\lambda_j,u_k,t_p,t_d)~\delta T(u_k,t_p) ~.
\end{equation}
Note that there is no net lag as in the lamppost model.  Our model does not assume a particular physical mechanism that is driving the temperature fluctuations and creating the radius-dependent lag in RM models.  Evidence for the lamppost model would be waves moving radially outward in $\delta T(R,t)$ with velocity equal to the speed of light.


\subsection{The linear algebra of modeling the data}

In Equation~\ref{eq:final_matrix}, $\delta F$ is a vector with length equal to the number of data points in our lightcurves, and $\delta T$ is a vector with length equal to the product of the dimensions of our temperature fluctuation map in radius $u$ and time $t_p$.  The $t_p$-array goes from the minimum and maximum dates of the array of observed times $t_d$ with constant grid spacing $\Delta t$.  One aspect of our model we will explore later in the paper is the dependence of the results on the time resolution determined by the dimension $N_t$ of our time grid.  The $u$-array goes from $u=1$ to $u=1000$ with equal logarithmic spacing.  The integrals that make up $W$ are evaluated as trapezoidal sums. To fit the lightcurves, we minimize the $\chi^2$ statistic
\begin{equation}
\chi^2  = \bigg(\frac{\delta F - W \delta T}{\sigma} \bigg)^T \cdot \bigg( \frac{\delta F - W \delta T}{\sigma} \bigg) 
\end{equation}
where $\sigma$ are the measured errors of $\delta F$.  It is convenient to work in terms of the modified arrays $\delta F_\sigma = \delta F / \sigma $ and $W_\sigma = W / \sigma$ such that
\begin{equation}
\chi^2  = (\delta F_\sigma - W_\sigma \delta T)^T \cdot (\delta F_\sigma - W_\sigma \delta T)  ~.
\end{equation}

While we are mapping two dimensions $(\lambda,t_d)$ into two dimensions $(u,t_p)$, this class of inversion problem generally leads to matrices whose inverses are undefined or ill-conditioned.  We address this by using linear regularization to stabilize the inversion and drive the solution to smoother temperature fluctuations.  Instead of minimizing $\chi^2$ alone, we minimize it along with additional terms that minimize the scale of the temperature fluctuations as well as the differences in the temperature fluctuations of adjacent radius and time bins. We define the first smoothing term as
\begin{equation}
H_0 = \xi_0 \bigg(\frac{\delta T}{T}\bigg)^T I \frac{\delta T}{T} \sim \xi_0 \bigg(\frac{\delta T}{T}\bigg)^2  
\end{equation}
where $I$ is the identity matrix and $\xi_0$ sets the strength of the smoothing.  This acts to minimize the amplitude of fractional temperature fluctuations.  Next, we also introduce smoothing along the model radial ($k$) and time ($l$) dimensions as 
\begin{equation}
\begin{split}
H_k = & ~\xi_k \bigg(\frac{\delta T}{T}\bigg)^T {D_k} \frac{\delta T}{T} \sim \xi_k \frac{(\delta T(u_{k+1}) - \delta T(u_k))^2}{T(u_{k+1}) T(u_k)} ~,~ {\rm and} \\
H_l = & ~\xi_l \bigg(\frac{\delta T}{T}\bigg)^T {D_l} \frac{\delta T}{T} \sim \xi_l \bigg(\frac{\delta T(t_{l+1}) - \delta T(t_l)}{T} \bigg)^2 \\
\end{split}
\end{equation}
where $D_k$ and $D_l$ are the first difference matrices for the model radial and time dimensions, respectively.  For the radial smoothing $H_k$, the fractional temperature fluctuation is calculated with respect to the geometric mean of the adjacent radius bins, hence the two different temperatures ($T(u_k)$ and $T(u_{k+1})$) in the equation.  It is convenient to work in terms of modified matrices $I_T = I/T^2$,  $D_{kT} = D_k/T^2$, and $D_{lT} = D_l/T^2$ such that 
\begin{equation}
\begin{split}
H_0 = & ~\xi_k (\delta T)^T I_T \delta T ~,\\
H_k = & ~\xi_k (\delta T)^T D_{kT} \delta T ~,~{\rm and}\\
H_l = & ~\xi_l (\delta T)^T D_{lT} \delta T  ~.\\
\end{split}
\end{equation}
We can now define the total linearization term
\begin{equation}
H = H_0 + H_k + H_l ~.
\end{equation}
We find that there are no interesting effects in scaling the three $\xi$'s separately, so we scale each smoothing matrix by the same $\xi = \xi_0= \xi_k = \xi_l$.  If we minimize $\chi^2 + H$  with respect to $\delta T$, we find that 
\begin{equation} \label{eq:final_linear}
\delta T = \Big[W_\sigma^T W_\sigma + \xi(I_T + D_{kT} + D_{lT})\Big]^{-1} W^T_\sigma \delta F_\sigma ~.
\end{equation}

The effects of changing $\xi$ will be one of our primary considerations.  Too much smoothing (high $\xi$), and the $\chi^2$ is poor.  Too little (low $\xi$), and $\delta T$ will show small scale structures with unphysically large $\delta T/T$. The standard approach is to choose $\xi$ such that the $\chi^2$ per data point $\chi^2/N_d \sim 1$. 

The lightcurves are in terms of the absolute flux $F$ rather than $\delta F$, and the mean flux of the lightcurve can differ from the model either because of contamination in the data (e.g., flux from the host galaxy) or problems with the steady-state disc model (e.g., wrong $\eta$ or $\lambda_{\rm Edd}$, or problems with the Shakura--Sunyaev model).  We are only interested in the patterns in $\delta T$ and not the mean or absolute fluxes, so we include an extra parameter for each lightcurve to remove any mean offsets between the data and the model.


\subsection{Filter kernels}

\begin{figure}
\includegraphics[width=\linewidth]{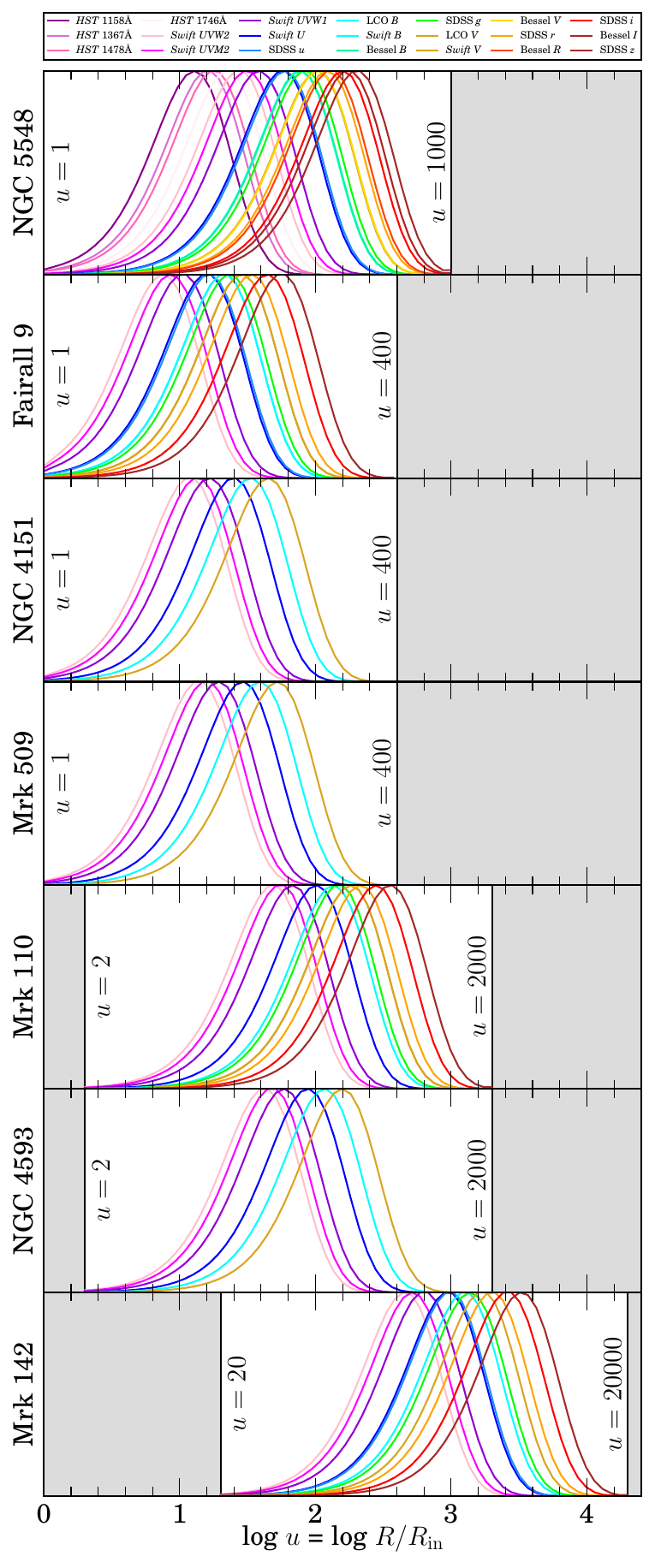}
\caption{Kernel contributions to various filters as a function of disc radius for our sample of AGNs.  Each kernel is normalized to be unity at peak.  Regions in $u$-space that are greyed out were not included in the analysis of a given AGN: for example, we only consider $u=1$ to $u=1000$ for NGC~5548 and $u=20$ to $u=20000$ for Mrk~142.  The kernels created by the blackbody (Eq.~\ref{eq:dfdt}) are so broad that there is no need to include the finite width of the filter.}
\label{fig:kernels}
\end{figure}

Because we have many filters in our lightcurve data, we must consider what range of radii (with different temperatures) will contribute to the flux in a given filter.  In Figure~\ref{fig:kernels}, we show the normalized flux contribution to each filter derived from Equation~\ref{eq:dfdt} as a function of radius for the AGNs we will model.  In Table~\ref{tab:agns}, we list the physical parameters of the AGN.  Because there are differing $M_{\rm BH}$ and $\lambda_{\rm Edd}$ values used to define the steady state temperature of the discs, the effective contribution of the radii to a given filter depends on the object.  For example, NGC~4151 and NGC~4593 were observed with the same filters, but the radii probed by the filters are very different because of the differing physical parameters.  Note that the filters must have significantly different wavelengths or else the signals will be strongly correlated since they probe similar radii in the disc.

\begin{table}
\centering
\caption{Physical parameters for the AGNs in our analysis}
\begin{tabular}{lccccc} \toprule
\multirow{2}{*}{AGN} & \multirow{2}{*}{$z$} & $D_L$ & $\log M_{\rm BH}$ & \multirow{2}{*}{$\lambda_{\rm Edd}$} & \multirow{2}{*}{Refs} \\
 &  & [Mpc] & $[\msun]$ & & \\ \midrule
NGC~5548  & 0.017 & 75   & 7.505 & 0.1 & 1 \\ 
Fairall~9 & 0.047 & 210  & 8.406 & 0.02 & 2,3 \\ 
Mrk~142   & 0.045 & 202  & 6.230 & 25 & 4,5 \\
Mrk~110   & 0.004 & 150  & 7.398 & 0.4 & 6 \\
NGC~4151  & 0.003 & 19.1 & 7.602 & 0.01 & 7 \\
NGC~4593  & 0.008 & 36   & 6.880 & 0.08 & 7 \\
Mrk~509   & 0.034 & 151  & 8.050 & 0.05 & 7 \\
\bottomrule
\end{tabular}
\begin{flushleft} \textit{Notes}: See Section~\ref{sec:agndata} for details on certain parameters.  References: (1) \citet{starkey17}; (2) \citet{vasudevan09}; (3) \citet{hernandez20}; (4) \citet{cackett20}; (5) \citet{li18}; (6) \citet{vincentelli21}; (7) \citet{edelson19}.  
\end{flushleft}
\label{tab:agns}
\end{table}

The wavelengths $\lambda$ that we will use are the effective central wavelengths of the filters for the lightcurves.  Including the finite widths of the filters would make the kernels in Figure~\ref{fig:kernels} slightly wider, but it would be a very small effect compared to the intrinsic width caused by the temperature profile of the disc and the structure of the blackbody kernel (Eq.~\ref{eq:dfdt}).  Furthermore, because these filter kernels are so wide, the dimension $N_u$ of our model grid in $u$-space is not a critical parameter, and so we used $N_u = 50$.  For NGC~4151, Mrk~509, and Fairall~9, a smaller range of $u$ values were probed by the filters (see Fig.~\ref{fig:kernels}), but rather than change the resolution of our $u$-space array, we trim off the parts of the grid with $u > 400$.


\subsection{The AGN and their data} \label{sec:agndata}

Table~\ref{tab:agns} lists the systems we will model along with their redshifts $z$, luminosity distances $D_L$, SMBH masses $M_{\rm BH}$, and Eddington ratios $\lambda_{\rm Edd}$.  

We carry out the most extensive analysis on the data for NGC~5548 from the AGN Space Telescope and Optical Reverberation Mapping (AGN STORM, \citealt{derosa15,edelson15,fausnaugh16,starkey17}).  These data consist of photometry using the \textit{Hubble Space Telescope} (\hst), \textit{Neil Gehrels Swift Observatory} (\swift, \citealt{gehrels04,roming05}), and various ground-based observatories.  The effective wavelengths stretch from 1158~\AA~to 8897~\AA. We also analyse Fairall~9 \citep{hernandez20}, Mrk~142 \citep{cackett20}, Mrk~110 \citep{vincentelli21}, NGC~4151 \citep{edelson17,edelson19}, NGC~4593 \citep{edelson19}, and Mrk~509 \citep{edelson19}.  The lightcurves of half of these AGNs (NGC~4151, NGC~4593, and Mrk~509) used only \swift, but the other half (Fairall~9, Mrk~142, and Mrk~110) were supplemented with ground-based observations in various filters (see Fig.~\ref{fig:kernels}).   

Sometimes we found that the reported errors for these data were too small.  We tried to correct for this using a ``triplet test'', where we fit three adjacent points in a lightcurve in a given filter $\lambda_j$ with a line.  The $\chi^2$ for each set of three points should be 1 if the errors are reasonable, so we calculated the offset $\sigma(\lambda_j,t_d)$ that when added to the errors in quadrature made $\chi^2=1$.  If this offset was negative, we set it to 0, as we did not want to decrease the errors.  We then computed the mean $\sigma(\lambda_j)$ and added this as a systematic increase to the errors in a given filter.  For many objects/filters, this correction was small.  

In Table~\ref{tab:agns}, we provide the references for $D_L$, $M_{\rm BH}$, and $\lambda_{\rm Edd}$.  Most of these values are consistent with those used in the lightcurve data papers, but some values had to be taken from other sources.  For Fairall~9, $\lambda_{\rm Edd} = 0.02$ is based on measurements by \citet{vasudevan09}, who derived the value from the X-ray SED of the source. For Mrk~142, we use the $\dot{m} = \lambda_{\rm Edd} / \eta = 250$ from \citet{li18}.  Since our Equation~\ref{eq:temp_in} assumes $\eta = 0.1$, we use $\lambda_{\rm Edd} = 25$.  When not provided in the papers listed in Table~\ref{tab:agns}, $D_L$ values were calculated using the redshifts and assuming conventional cosmological parameters (flat universe, $h = 0.696$, and $\Omega_{\rm m} = 0.286$, \citealt{wright06}).  

The $\delta F$, $\lambda_j$, and $t_d$ values used in our equations are redshift-corrected before we fit the data.  However, when we plot the model temperature fluctuation patterns, the times are presented in observed (i.e., dilated) times, not host-rest-frame times, so that they can be easily compared to the lightcurves. 

Increasing the inclination $i$ has two effects on the model: decreasing the observed flux relative to the luminosity (see Eq.~\ref{eq:flux}) and increasing the differential light travel time (see Eq.~\ref{eq:t0}).  Since we are only concerned about fractional changes, the first effect has no impact on the model. The second effect is usually only important for large $u$ values, as even at $i=90\degr$ the light travel time is very small for $u<100$ in most of our systems.  We choose an arbitrary $i = 30\degr$ for all of our models, since we found no noticeable effects from varying it.


\section{Testing our model}\label{sec:testing}

\begin{figure*}
\includegraphics[width=\linewidth]{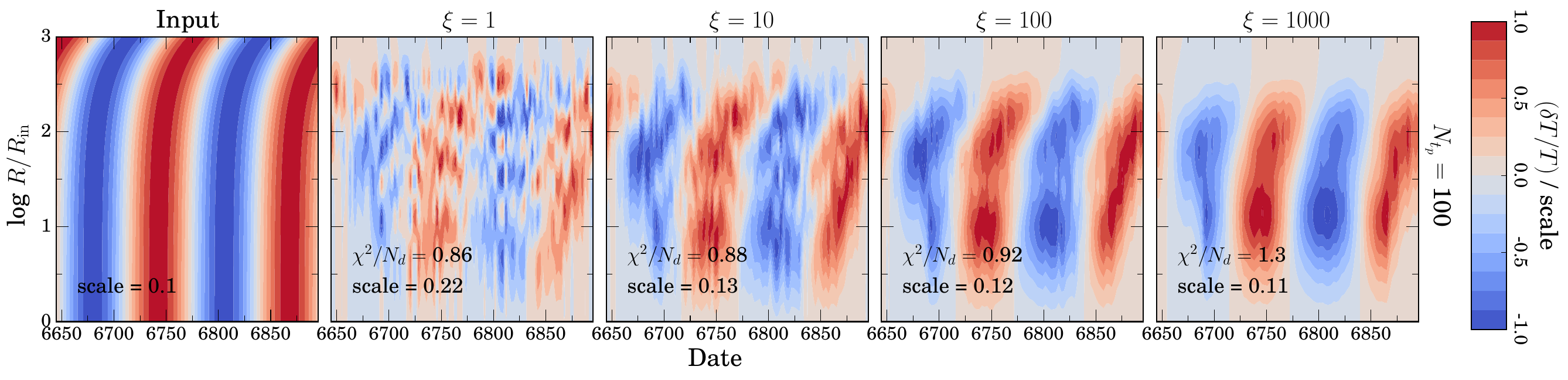}
\caption{Input and reconstructed temperature maps for the \textit{outgo} (fast outgoing) test pattern with $N_t = 100$ and $\xi = 1,10,100,1000$.   Each inversion gives the $\chi^2$ per data point and the scale of $|\delta T/T|$ for the colorbars.  The color scale is set as the 99th percentile of the range of $|\delta T/T|$ values.  A lamppost signal will look like these \textit{outgo} reconstructions but with patterns nearly vertical because a lamppost signal moves at $c$.  The vertical/temporal structure near 6700~days, which is more prominent in Fig.~\ref{fig:ingo_panel}, is due to the ``sudden'' addition of the \hst~and \swift~data.}
\label{fig:outgo_panel}
\end{figure*}
\begin{figure*}
\includegraphics[width=\linewidth]{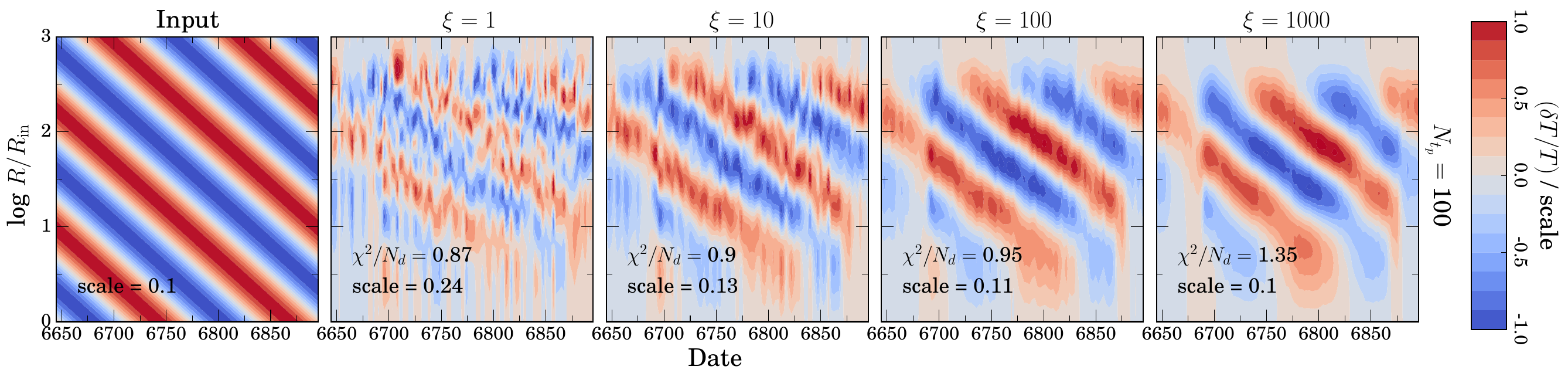}
\caption{Input and reconstructed temperature maps for the \textit{ingo} (slow ingoing) test pattern with $N_t = 100$ and $\xi = 1,10,100,1000$. }
\label{fig:ingo_panel}
\end{figure*}
\begin{figure*}
\includegraphics[width=\linewidth]{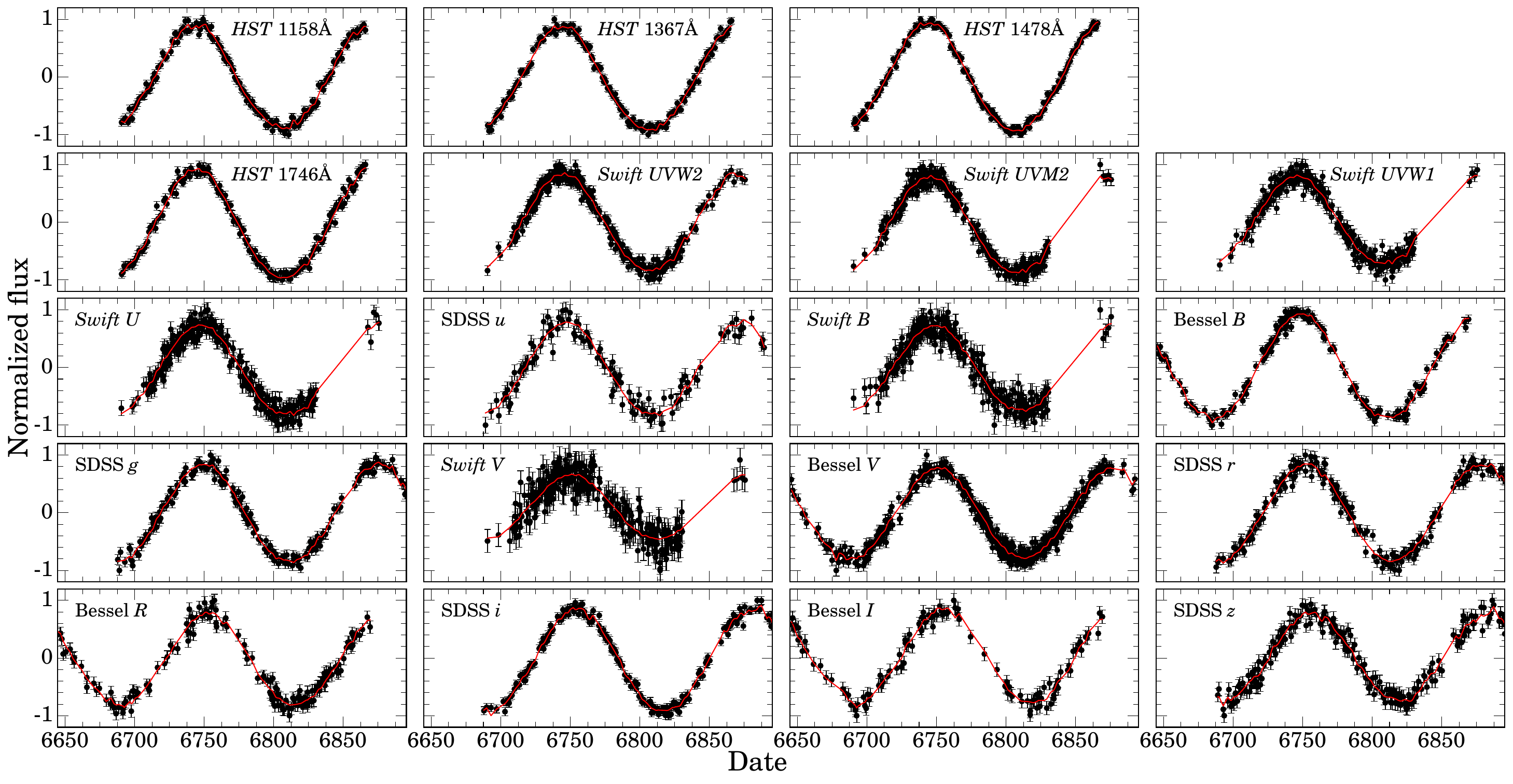}
\caption{Input (black) and model (red) lightcurves for the \textit{outgo} (fast outgoing) test pattern with $N_t = 100$ and $\xi=10$.  The lightcurves are normalized so that the minimum and maximum observed flux values are --1 and +1, respectively.}
\label{fig:outgo_lc}
\end{figure*}

The next step is to illustrate that the model works on test problems.  In this section, we test two $\delta T$ patterns: \textit{outgo}, where the temperature fluctuations move out from the center of the disc at $0.1c$, and \textit{ingo}, where the temperature fluctuations move inward at some arbitrary speed that decreases as it moves inward.  The patterns are repeated in time with a period of 20~d that is similar to the typical variability time-scales of the lightcurves.  The \textit{outgo} pattern is meant to mimic the lamppost model, albeit slowed down to accentuate the time-lag at larger radii.   The \textit{ingo} pattern is set up such that $v \propto u$ and $v \sim 0.025c$ at $u=100$.  A key difference between the \textit{outgo} and \textit{ingo} patterns -- in addition to the difference in their propagation speeds -- is that the former produces lightcurves with blue variability leading red, as generally observed in RM studies, whereas the latter has red variability leading blue.  For physical parameters, filters, and observed times, we use those of NGC~5548.  In Appendix~\ref{sec:appendix}, we show tests for two additional temperature fluctuation patterns.

For each test pattern, we generate the temperature fluctuations and set the maximum and minimum values of the fluctuations to be 10~per~cent of the disc temperature at the respective radius.  We then add those temperature fluctuations to the disc temperature profile, and directly calculate the lightcurves with these fluctuations using Equation~\ref{eq:flux} with $T = T_0 + \delta T(R,t)$.  By generating the lightcurves with the non-linear dependence on $\delta T$ of Equation~\ref{eq:flux}, we can evaluate the effects of reconstructing them using the linearized equations.  The time-delay effects are also included in this integration. The errors of these synthetic fluxes are chosen such that the ratios of the errors to the RMS flux variations are the same for the test data as for the mean errors of the real data for NGC~5548.  We also add a Gaussian realization of this noise level to each synthetic data point.  We then reconstruct $\delta T$ using the linearized equations as a function of the smoothing parameter $\xi$.  The input and reconstructed temperature maps for the \textit{outgo} and \textit{ingo} patterns are shown in Figures~\ref{fig:outgo_panel} and \ref{fig:ingo_panel}, respectively.  As can be seen, we are able to accurately reconstruct the input temperature fluctuations despite linearizing the reconstruction and the presence of noise. For the \textit{outgo} pattern, we also show the synthetic and reconstructed lightcurves in Figure~\ref{fig:outgo_lc} using $\xi = 10$, and we can see that our model accurately reconstructs the synthetic lightcurves.  One can see that the lightcurves in each filter resemble each other, but with a blue-leading-red lag of a few days, similar the lamppost signal observed in real AGN lightcurves.  

Looking closer at Figures~\ref{fig:outgo_panel} and \ref{fig:ingo_panel}, we note that the goodness of fit improves with less smoothing, but the temperature fluctuations have higher amplitude and more structure than the input fluctuations.  For example, the peak amplitudes of the $\xi = 1$ case in Figures~\ref{fig:outgo_panel} and \ref{fig:ingo_panel} are larger than the input fluctuations by a factor of 2 or more.  This is the standard issue for these types of inversion problems.  The better the fit to the data, the amplitudes of the inverted temperature increase and the physical scales decrease.  The linear inversions closely match the non-linear inputs provided we use the smoothing to keep the fractional amplitude below 10--20~per~cent.  As the fractional amplitude increases, the linear and non-linear lightcurves generated by the fluctuations diverge.  If they become too large, the linearized model can even produce unphysical negative temperatures and fluxes.  This also means that our linearized model is not suitable for modeling changing look (e.g., \citealt{dexter19}) or hypervariable \citep{rumbaugh18} AGNs, where the flux variations are large.

\begin{figure*}
\includegraphics[width=\linewidth]{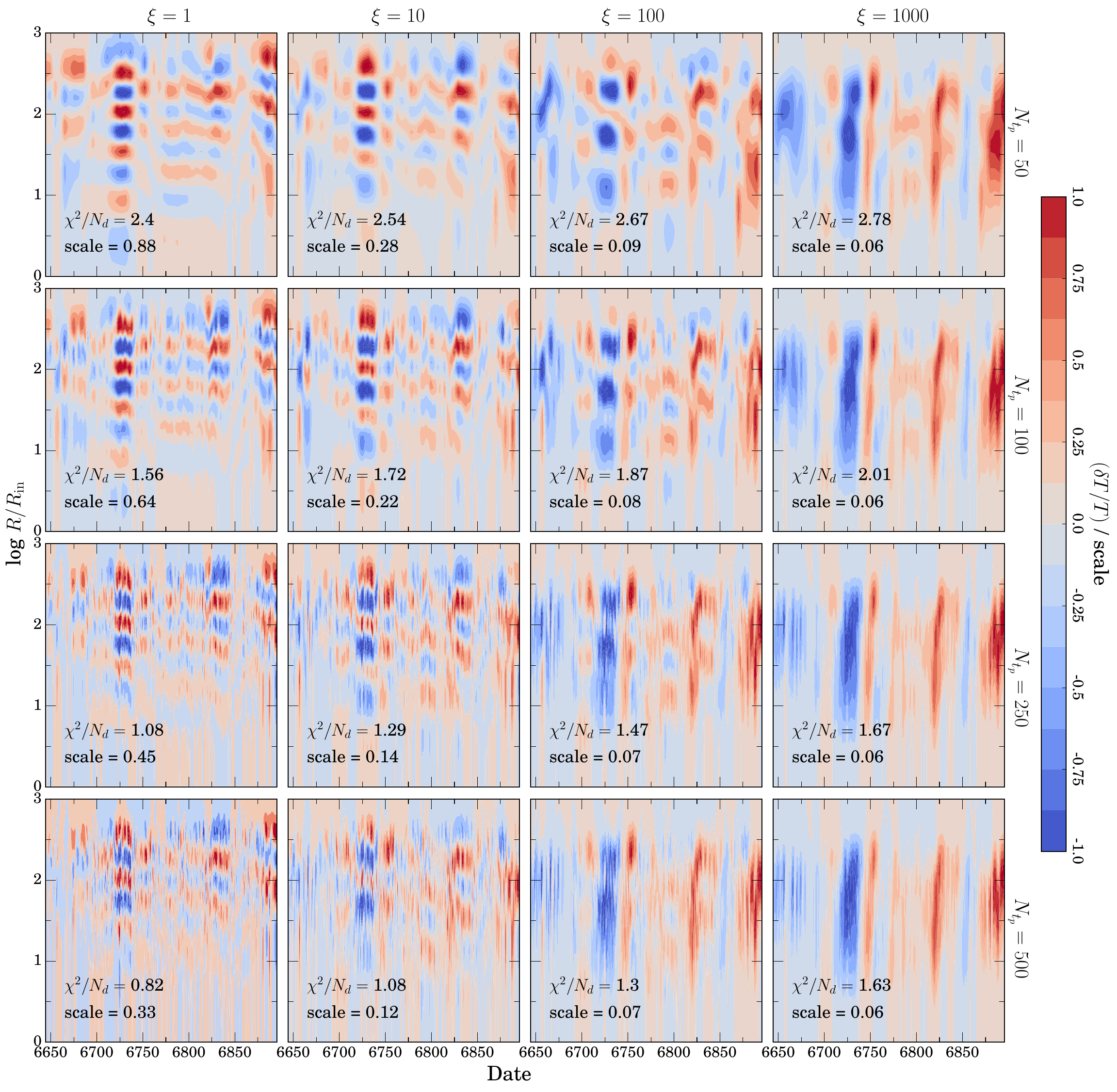}
\caption{Temperature maps of NGC~5548 with temporal dimensions of $N_{t} = 50, 100, 250, 500$ (rows) and smoothing factors of $\xi = 1,10,100,1000$ (columns).  Maps with high $\xi$ look lamppost-like, whereas those with low and moderate $\xi$ show significant deviations from the lamppost model in the form of radial and ingoing structures.  Additionally, maps more densely sampled in time (higher $N_{t}$) show more small-scale structure, but still resemble those with lower $N_{t}$.}
\label{fig:ngc5548_panel}
\end{figure*}
\begin{figure*}
\includegraphics[width=\linewidth]{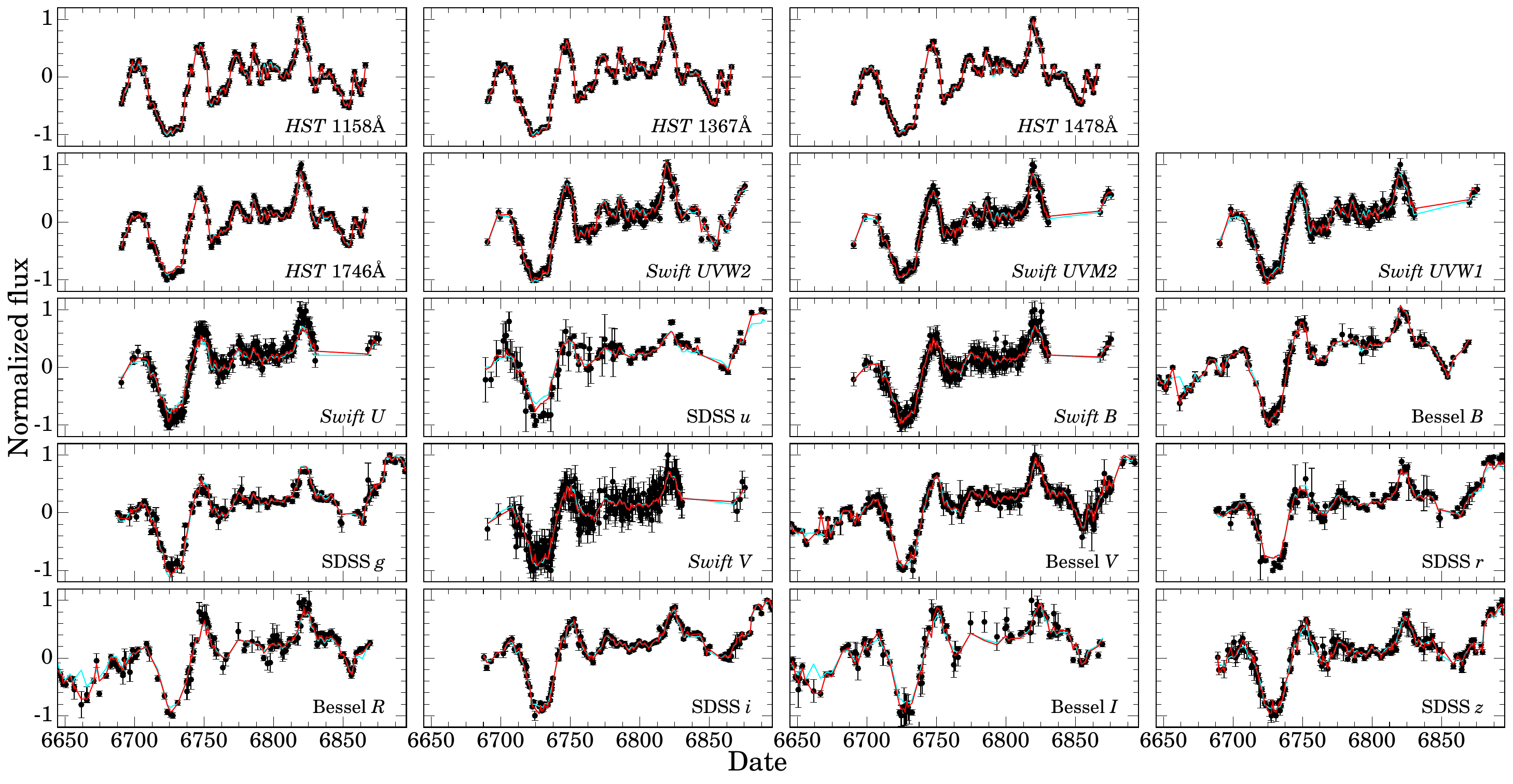}
\caption{Observed (black) and model lightcurves of NGC~5548 with $N_t = 250$ and $\xi=1$ (red, $\chi^2/N_d = 1.08$) and $\xi=1000$ (cyan, $\chi^2/N_d = 1.67$).}
\label{fig:ngc5548_lc}
\end{figure*}


\section{NGC~5548}\label{sec:ngc5548}

\begin{figure*}
\includegraphics[width=\linewidth]{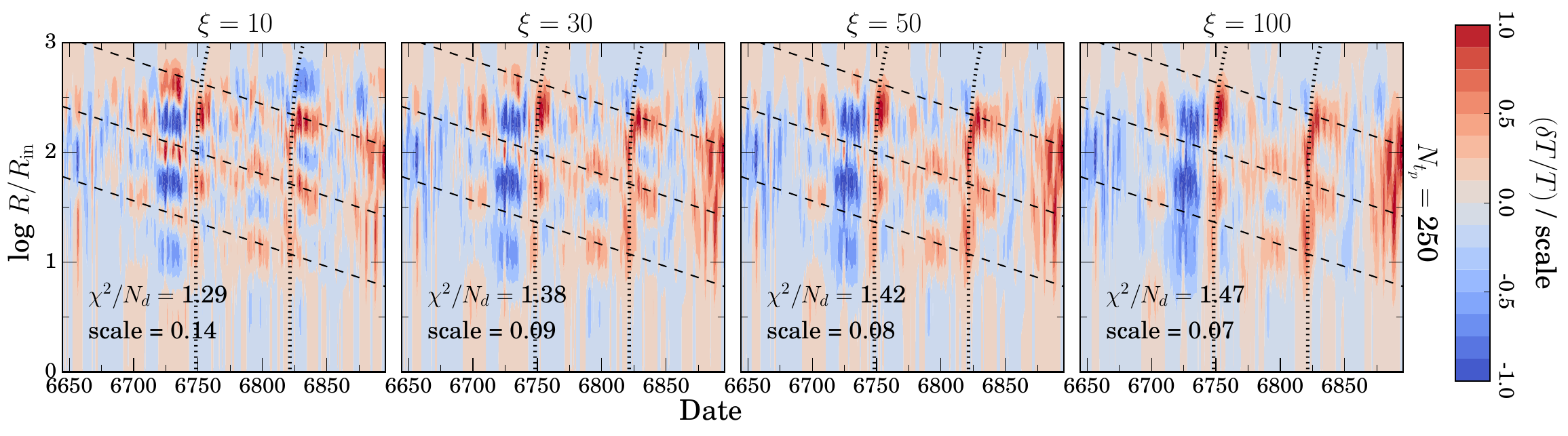}
\caption{Temperature maps of NGC~5548 with $N_t = 250$ and $\xi = 10,30,50,100$.  Here we highlight the patterns that appear prominent in the temperature maps.  The dashed lines correspond to ingoing patterns with velocity $v \propto u$.  Each line is separated by 160~d, and the velocity at $u=100$ is $\simeq 0.01c$.  The nearly-vertical dotted lines around 6700 and 6825~days correspond to lamppost-like outgoing patterns with $v=c$.}
\label{fig:ngc5548_more}
\end{figure*}

We study NGC~5548 first and in more detail to illustrate the dependence of the results on various model choices like array dimensions and smoothing.  As discussed earlier, we used $N_u=50$ for all of our solutions because the large width of the filter kernels (see Fig.~\ref{fig:kernels}) means that changes in the radial griding have little effect.

We explored but do not show the effects of changing inclination $i$ and Eddington ratio $\lambda_{\rm Edd}$.  As we discussed earlier, the former has little effect on the solutions, so we assume an arbitrary $i = 30\degr$.  We assume $\lambda_{\rm Edd} = 0.1$, but \citet{vasudevan09} measure $\lambda_{\rm Edd} = 0.0236$, and other papers use different other $\lambda_{\rm Edd}$ (e.g., $\lambda_{\rm Edd} = 0.05$, \citealt{edelson19}).  Changing $\lambda_{\rm Edd}$ changes the radii probed by the filters, but the structures and patterns seen in the fluctuations do not change significantly.  Since we are mostly interested in these patterns and not in their precise physical locations, we do not vary $\lambda_{\rm Edd}$ in our solutions and fix it to $\lambda_{\rm Edd} = 0.1$. Instead, the primary variables affecting the results are the smoothing parameter $\xi$ and the time resolution set by the number of temporal zones $N_t$.  The primary issue for $N_t$ is that it sets the overall matrix size, ultimately leading to issues with execution time and memory requirements.  

In Figure \ref{fig:ngc5548_panel}, we show the temperature models constructed using $N_t$ ranging from 50 to 500 and $\xi$ ranging from 1 to 1000.  In each panel of Figure~\ref{fig:ngc5548_panel}, the maps have more radial structure and definition for dates between 6700--6850~days.  This corresponds to the range of dates where \hst~and \swift~data were available.  Elsewhere, only ground-based data were available, and thus the model has less information and hence a less well-defined inversion.

In Figure \ref{fig:ngc5548_lc}, we show the observed lightcurves and the model lightcurves for $N_t=250$ and $\xi=1$ and $\xi = 1000$.  While there is an increasing $\chi^2$ for higher $\xi$, the differences between the reconstructed lightcurves of various smoothings are very subtle.  The two smoothings shown in Figure~\ref{fig:ngc5548_lc} have a difference in $\chi^2/N_d$ of 0.59, yet the only obviously visible differences between the lightcurves are where the $\xi=1$ lightcurve is better at replicating some early-time features in the Bessel~$B$, $V$, and $R$ filters.  As such, for other AGNs, we only show the reconstructed lightcurves for $\xi = 10$.

One particular feature is that both of the reconstructed lightcurves fail to accurately reproduce the brightest and faintest parts of the \textit{Swift U} (3494~\AA) and SDSS~\textit{u} (3590~\AA) lightcurves, with larger discrepancies than for any other filter.  This is most likely due to contamination from emission lines, namely the Balmer jump around 3645~\AA~\citep{korista01,lawther18}.  In disc RM lag measurements, these filters also show anomalous lags, which are attributed to the same emission line contamination (e.g., \citealt{,fausnaugh16,edelson19}).  Removing these filters would improve our fits, but it would also create a gap in the radii probed by the model (see Fig.~\ref{fig:kernels}).  Keeping them does not seem to drive any anomalous features in the reconstruction.  

As shown by Figure~\ref{fig:ngc5548_panel}, the $\chi^2$ decreases with decreasing $\xi$ and increasing $N_t$.  The scale of the fluctuations decreases with increasing $\xi$ and increasing $N_t$.  The relation between the scale of fluctuations and $\xi$ is expected because increasing $\xi$ forces the fluctuations to be smaller, but the relation between the scale and $N_t$ is more complicated.  A possible explanation is that, with increasing $N_t$, the largest fluctuations can be spread out over multiple time intervals rather than concentrated in a single interval of $t_p$ space.  With spread-out, scaled-down fluctuations, the weighting of the smoothing term goes down and thus the $\chi^2$ is better minimized.  For $\xi = 1$, the fractional scale of fluctuations is very large, to the point of being unphysical.  As we discussed in Section~\ref{sec:testing}, the reconstructed models for the test problems with >20~per~cent fluctuations would provide poor fits to the lightcurve when directly integrated with the blackbody equations (Eq.~\ref{eq:flux}) rather than the linear expansion (Eq.~\ref{eq:dfdt}).

Increasing $\xi$ has the general effect of making the model look more ``lamppost-like,'' in that the temperature fluctuations become nearly vertical (i.e., fast) outgoing waves with only small radial variations along the path of the wave.  This makes sense, as the observed lightcurves are dominated by a lamppost-like signal. One can look at the $\xi = 1000$ models in Figure~\ref{fig:ngc5548_panel} and the lamppost-like \textit{outgo} test models in Figure~\ref{fig:outgo_panel} and note the visual similarities.  If we were interested in better approximating the lamppost model using this method, we could change the smoothing model to heavily smooth along these outgoing signals and less so in the transverse direction.  Nevertheless, these lamppost-like inversions have higher $\chi^2$ values and are thus not as good at fitting the data.  

The most prominent and striking feature of these models is that there are coherent ingoing wave patterns in the temperature maps for $\xi=1,10,100$, very similar in appearance to the test \textit{ingo} model in Figure~\ref{fig:ingo_panel}.  For higher $\xi$ values, the ingoing wave patterns appear smeared out and are no longer discernible.  Additionally, when we look at increasing $N_t$ from 50 to 500, the general patterns still exist for a given $\xi$, but fluctuations on short time-scales ($\sim$1~d) become more prominent.  If we look at $\xi = 10$ with $N_t$ ranging from 50 to 250, the temperature fluctuations over time seem to repeat with increasing radius, going from negative to positive and back.  These radial patterns move inward, which in the map, appear to have the same ``speed''.  Because the scale of $u$ is logarithmic and time is linear, this means that the speed of the pattern is decreasing as it moves inwards with a velocity roughly following $v \propto u$.  This consequently means that the time-scale for radial changes $\tau \sim u/v$ is a constant.

\begin{figure*}
\includegraphics[width=\linewidth]{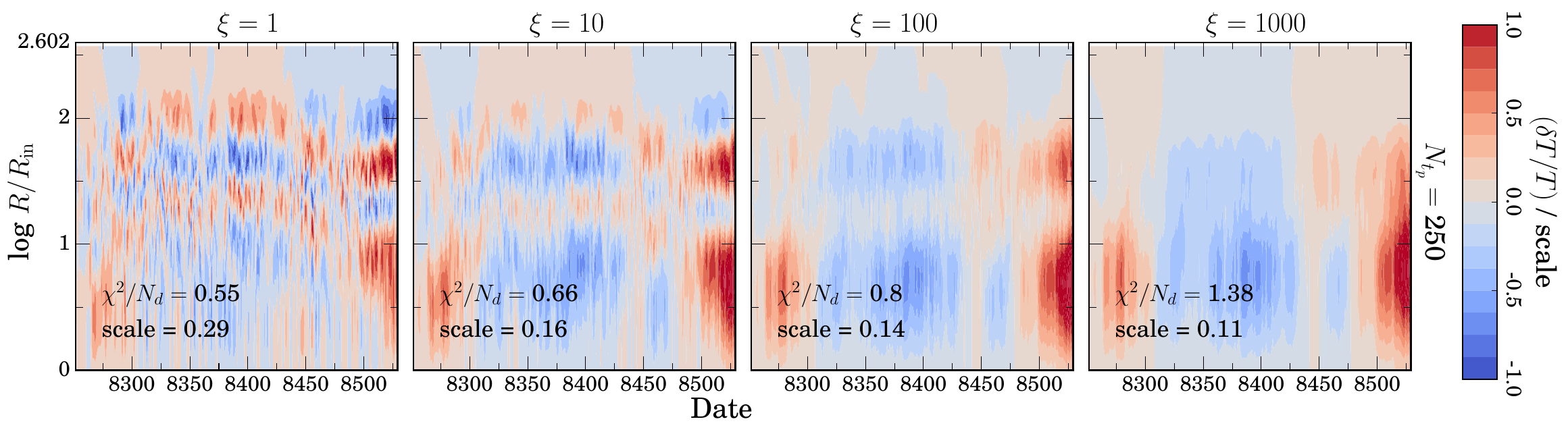}
\caption{Temperature maps of Fairall~9 with $N_t = 250$ and $\xi = 1,10,100,1000$.  Except at highest $\xi$, there is evidence for coherent radial structure which slowly moves radially outwards before moving inwards.}
\label{fig:fairall9_panel}
\end{figure*}
\begin{figure*}
\includegraphics[width=\linewidth]{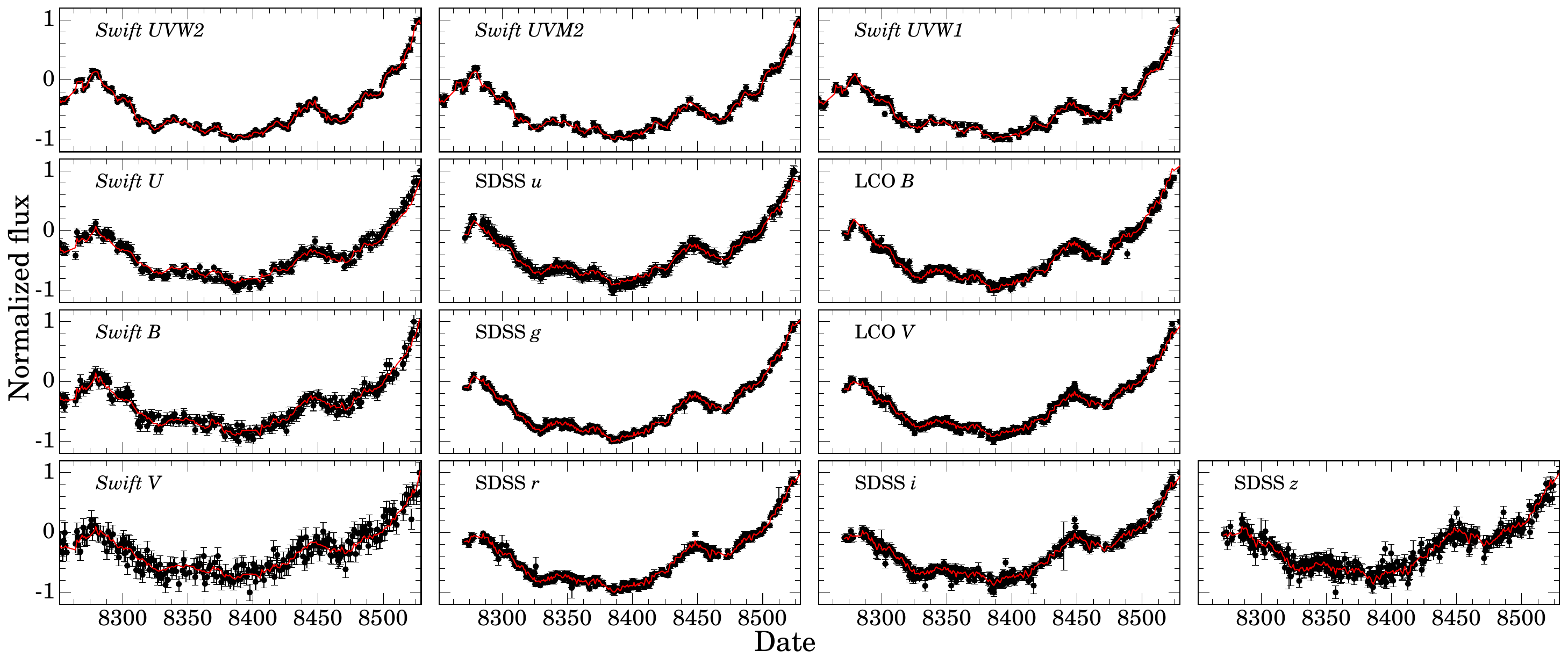}
\caption{Observed (black) and model (red) lightcurves of Fairall~9 with $N_t = 250$ and $\xi=10$.}
\label{fig:fairall9_lc}
\end{figure*}

In Figure \ref{fig:ngc5548_more}, we examine these ingoing patterns in greater detail.  We overlay dashed lines that track the repeated ingoing pattern with $v \propto u$.  We also include lines for a lamppost signal moving outward at the speed of light, and these closely track the more vertical structures.  The slopes and locations of these lines are chosen by eye and were not numerically derived.  The lines appear to track the general structure quite well, though there are more complex features, such as the dramatic change in the radial profile between 6725--6775~days corresponding to a similarly dramatic feature in the lightcurves (see Fig.~\ref{fig:ngc5548_lc}).  The dashed lines are all set to have the same velocity profile, with a velocity at $u=100$ of roughly 3000~km~s$^{-1}\simeq 0.01c$.  This is $0.24$ times the orbital velocity at this radius.  The lines are separated by $\tau =160$~d (157~d in host rest-frame) which is similar to the orbital time-scale at $u=100$ of $\tau_{\rm orb} = 168$~d.   However, the orbital time-scale increases with $u$ as $\tau_{\rm orb} \propto u^{3/2}$, whereas the time-scales for these ingoing patterns are constant, so similarity to the orbital time-scale only exists for a small range of $u$. In any case, the repeating banded patterns seen in our models for NGC~5548 are very interesting and, as we shall see later, share commonalities with the results for other AGNs.


\section{Analysis of other AGNs}\label{sec:others}

We repeat our analysis for Fairall~9, Mrk~142, Mrk~110, NGC~4151, NGC~4593, and Mrk~509.  In each temperature map for these AGNs, we only show the solutions for one optimized value of $N_t$, chosen based on the time resolution and range of the data.  We still show a range of $\xi$ solutions to show the effect of smoothing.  Because of the changing values of $\lambda_{\rm Edd}$ and $M_{\rm BH}$, the disc regions constrained by the data depends on the AGN, which is reflected in the vertical axes of the temperature maps and in Figure~\ref{fig:kernels}.  The logarithmic resolution of the $u$-array is the same for all of the AGNs.


\subsection{Fairall~9}

In Figure \ref{fig:fairall9_panel}, we show the temperature maps constructed using $N_t = 250$ and $\xi$ ranging from 1 to 1000.  In Figure \ref{fig:fairall9_lc}, we show the observed lightcurves and the model lightcurves constructed using $\xi=10$.  For $\xi \leq 10$, the temperature map pattern of Fairall~9 somewhat resembles its lightcurves: there are coherent radial temperature perturbations which move outwards until $\sim$8400 days and then moves back inwards; in the lightcurves, the flux drops and rises in a parabola-like curve around the same times.  The slowly-varying component of the lightcurves was noted in \citet{hernandez20}, who also note that this slow component has a lag signal opposite to that of a lamppost signal (i.e., red-leading-blue instead of blue-leading-red), but the physical implications of it were not explored in depth.  The time-scale of these radial changes is the full range of the observed epochs, around 300~d, which is of order the $\tau_{\rm orb}$ for intermediate radii (for Fairall~9, $u=40$ corresponds to $\tau_{\rm orb} = 339$~d). These patterns disappear above $\xi=100$, in favor of a more lamppost-like model, similar to what we see with NGC~5548 in Figure~\ref{fig:ngc5548_panel}.  Interestingly, even at the highest smoothing considered of $\xi=1000$, there is a radial structure between 8450--8475~days where below $u = 10$, the fluctuations are negative, and above it, they are positive. This should not exist in a pure lamppost model, especially for such a long time interval compared to the light travel time at these radii. 


\subsection{Mrk~142}

\begin{figure*}
\includegraphics[width=\linewidth]{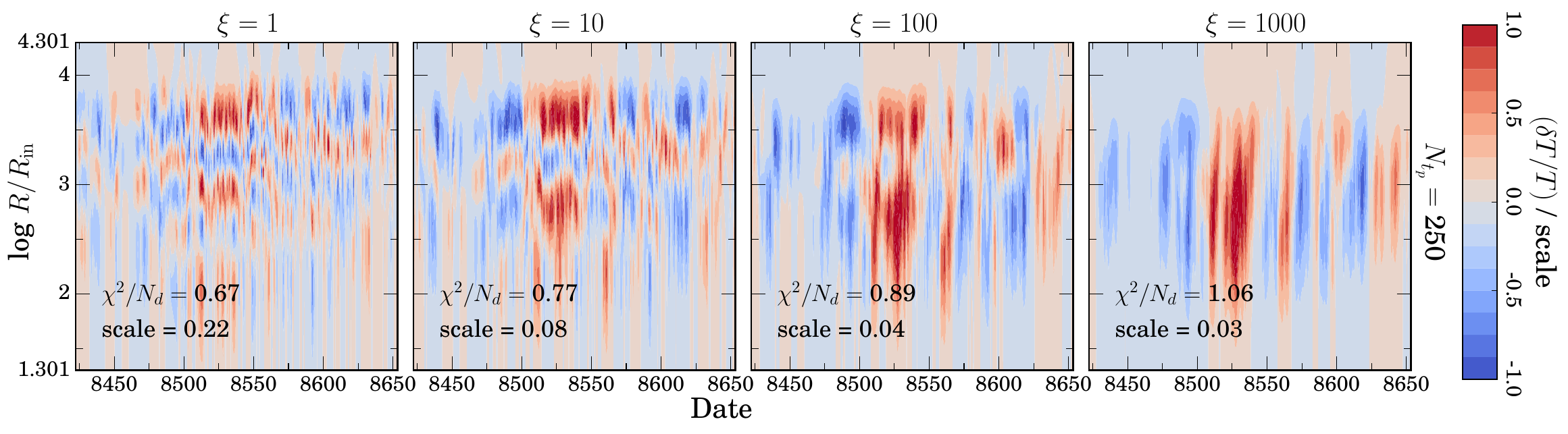}
\caption{Temperature maps of Mrk~142 with $N_t = 250$ and $\xi = 1,10,100,1000$.  There seem to be no coherent structures in these maps, but there are some radial structures which should not exist in a pure-lamppost model.}
\label{fig:mrk142_panel}
\end{figure*}
\begin{figure*}
\includegraphics[width=\linewidth]{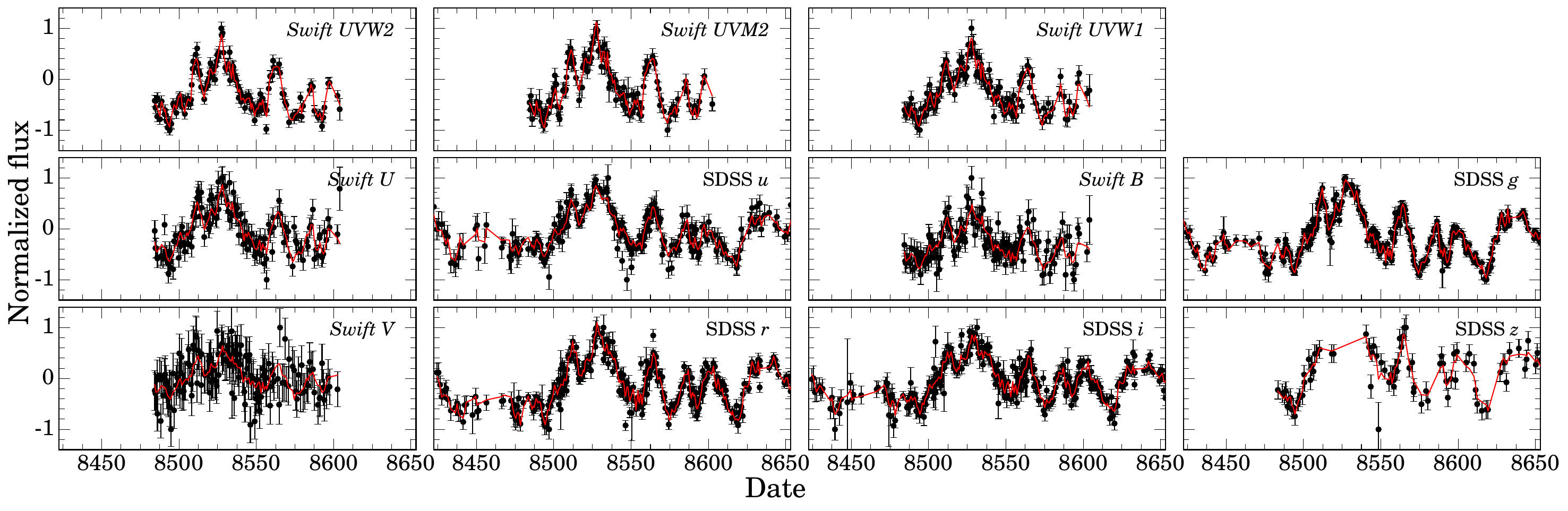}
\caption{Observed (black) and model (red) lightcurves of Mrk~142 with $N_t = 250$ and $\xi=10$.}
\label{fig:mrk142_lc}
\end{figure*}

In Figure \ref{fig:mrk142_panel}, we show the temperature maps constructed using $N_t = 250$ and $\xi$ ranging from 1 to 1000.  In Figure \ref{fig:mrk142_lc}, we show the observed lightcurves and the model lightcurves constructed using $\xi=10$.  Even at low $\xi$, there are no obvious coherent patterns or ingoing/outgoing waves in the temperature fluctuations, although there are prominent radial fluctuations which change with time.  Like the previous cases, these radial fluctuations disappear with high smoothing in favor of a lamppost-like pattern. It is worth noting that Mrk~142 is the only AGN in our sample that has a Eddington ratio estimate greater than unity.  Perhaps super-Eddington accretion is the reason it has more ``incoherent'' temperature fluctuations than the other sources.


\subsection{Mrk~110}

\begin{figure*}
\includegraphics[width=\linewidth]{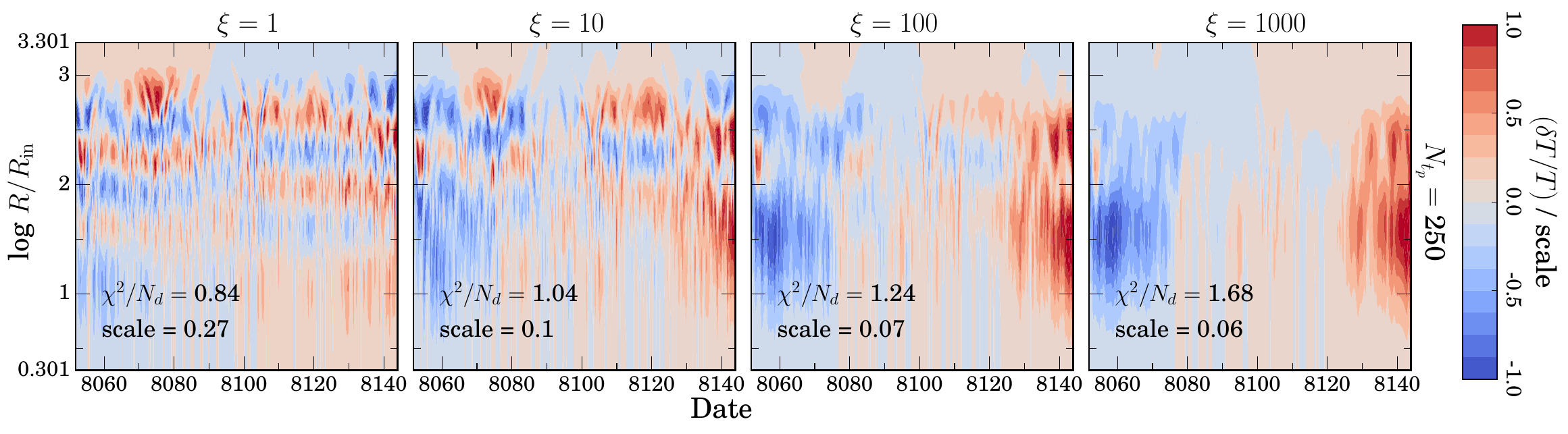}
\caption{Temperature maps of Mrk~110 with $N_t = 250$ and $\xi = 1,10,100,1000$. Like Fairall~9 (Fig.~\ref{fig:fairall9_panel}), there is evidence for coherent radial structures which slowly move radially outwards before moving inwards.}
\label{fig:mrk110_panel}
\end{figure*}
\begin{figure*}
\includegraphics[width=\linewidth]{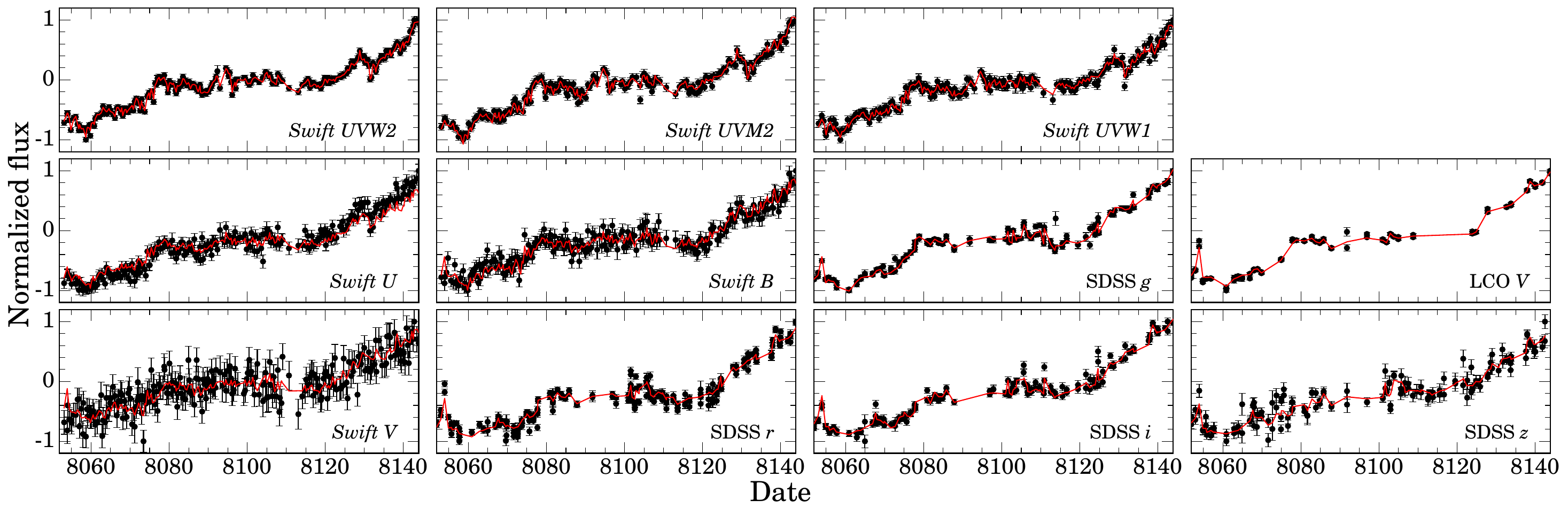}
\caption{Observed (black) and model (red) lightcurves of Mrk~110 with $N_t = 250$ and $\xi=10$.}
\label{fig:mrk110_lc}
\end{figure*}

\begin{figure*}
\includegraphics[width=\linewidth]{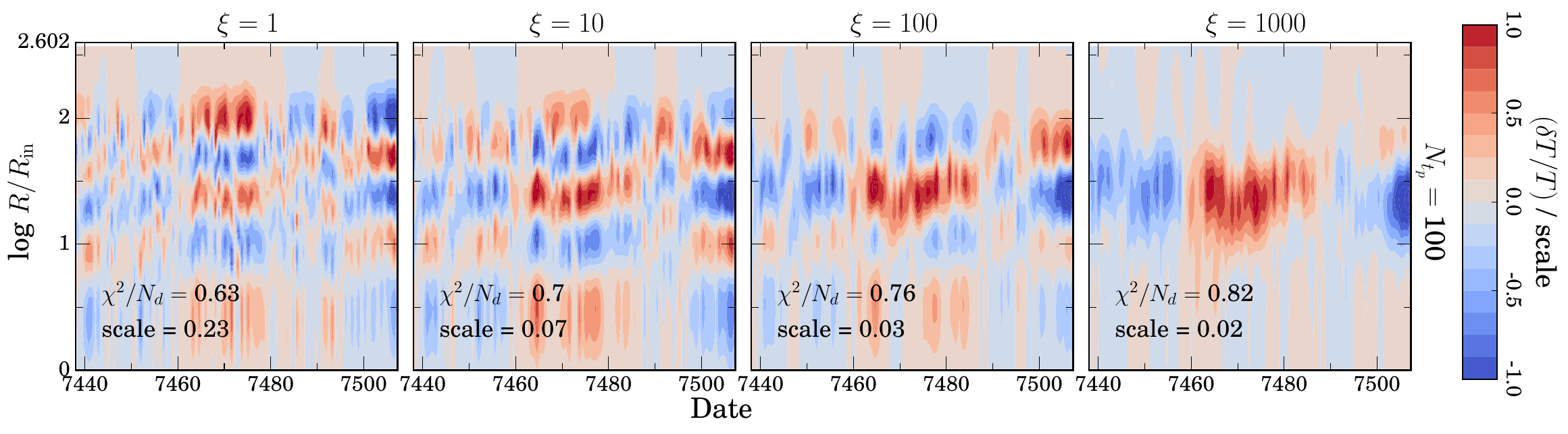}
\caption{Temperature maps of NGC~4151 with $N_t = 100$ and $\xi = 1,10,100,1000$.  Compared to other AGNs, the radial structures are stable to even the highest smoothing/$\xi$ values.}
\label{fig:ngc4151_panel}
\end{figure*}
\begin{figure*}
\includegraphics[width=\linewidth]{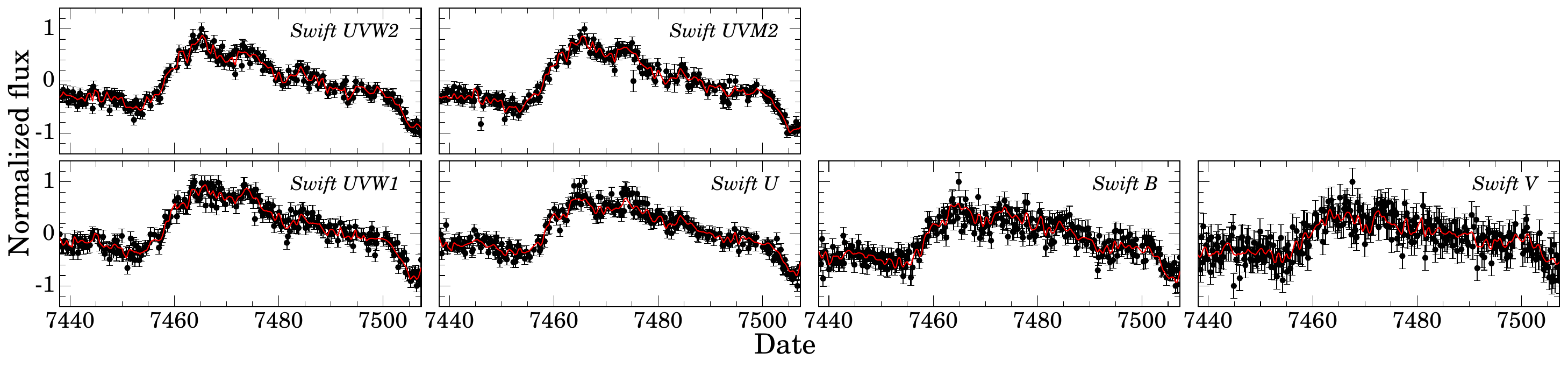}
\caption{Observed (black) and model (red) lightcurves of NGC~4151 with $N_t = 100$ and $\xi=10$.}
\label{fig:ngc4151_lc}
\end{figure*}

\begin{figure*}
\includegraphics[width=\linewidth]{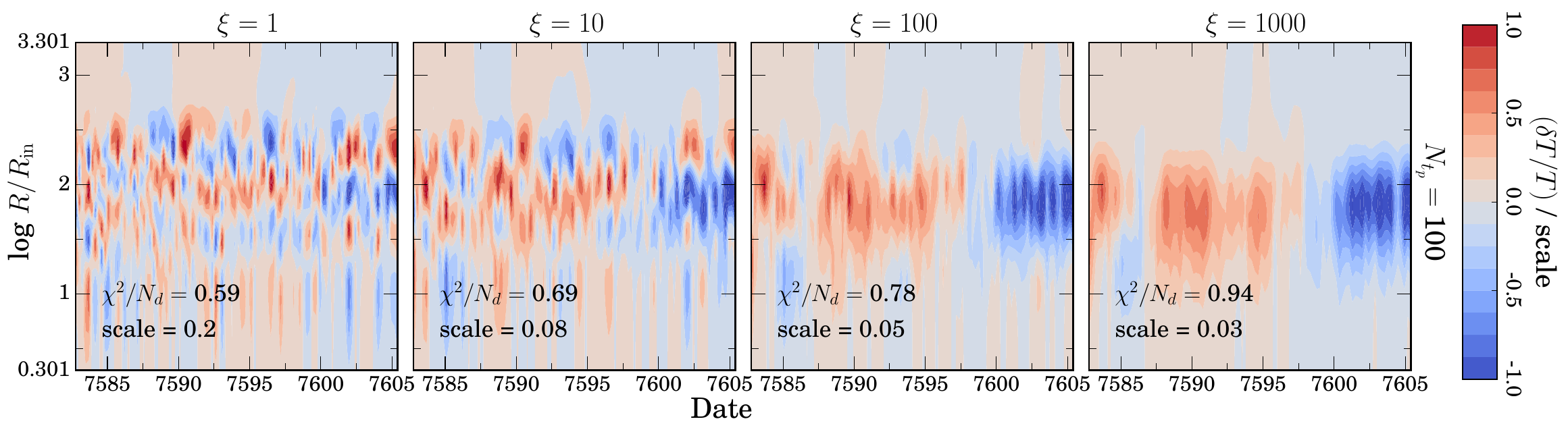}
\caption{Temperature maps of NGC~4593 with $N_t = 100$ and $\xi = 1,10,100,1000$.  There do not appear to be any obvious coherent patterns in any of these maps.}
\label{fig:ngc4593_panel}
\end{figure*}
\begin{figure*}
\includegraphics[width=\linewidth]{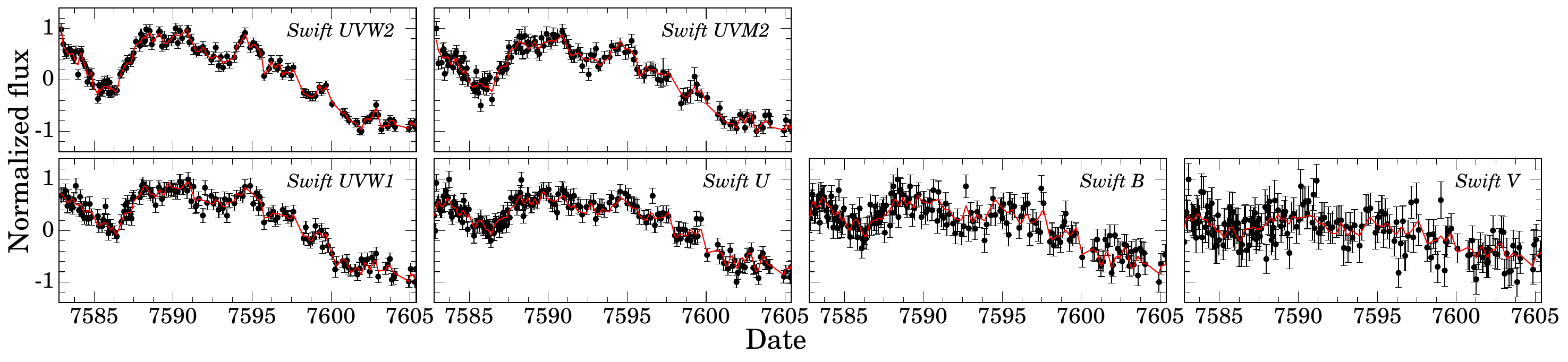}
\caption{Observed (black) and model (red) lightcurves of NGC~4593 with $N_t = 100$ and $\xi=10$.}
\label{fig:ngc4593_lc}
\end{figure*}

\begin{figure*}
\includegraphics[width=\linewidth]{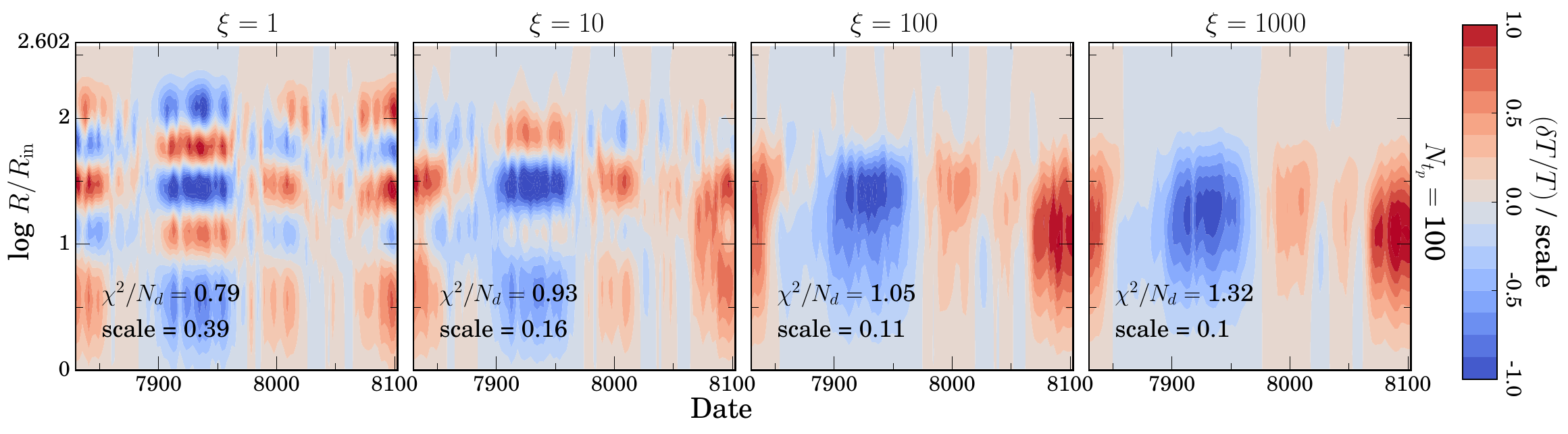}
\caption{Temperature maps of Mrk~509 with $N_t = 100$ and $\xi = 1,10,100,1000$.  In the $\xi=1,10$ cases, there is evidence for ingoing radial structure from 8000--8050~days, but it is smoothed out at higher $\xi$ values.}
\label{fig:mrk509_panel}
\end{figure*}
\begin{figure*}
\includegraphics[width=\linewidth]{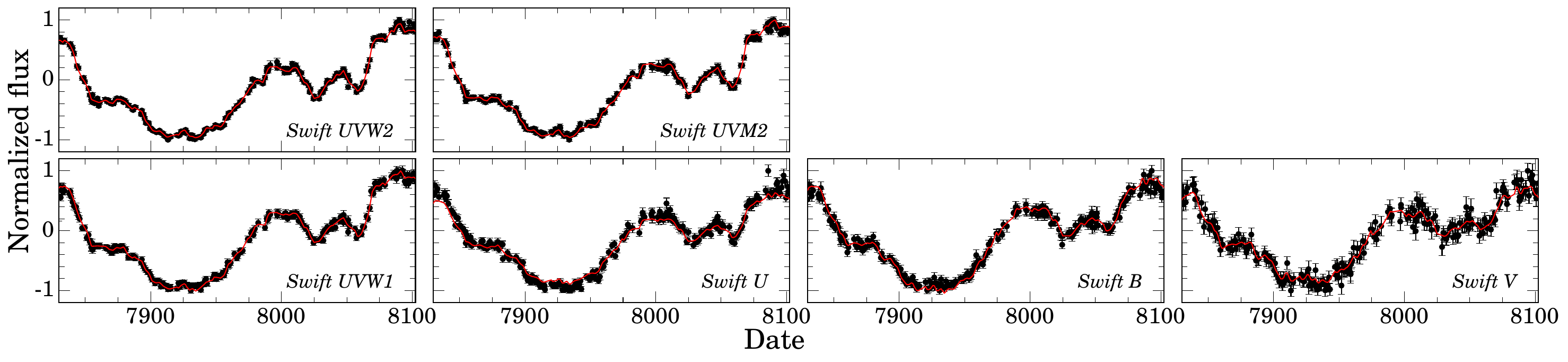}
\caption{Observed (black) and model (red) lightcurves of Mrk~509 with $N_t = 100$ and $\xi=10$.}
\label{fig:mrk509_lc}
\end{figure*}

In Figure \ref{fig:mrk110_panel}, we show the temperature maps constructed using $N_t = 100$ and $\xi$ ranging from 1 to 1000.  In Figure \ref{fig:mrk110_lc}, we show the observed lightcurves and the model lightcurves constructed using $\xi=10$.  Similar to NGC~5548 and Fairall~9, there is clear radial structure observed in the $\xi \leq 10$ models in the form of alternating positive and negative temperature fluctuations in radius.  These fluctuations move slowly outwards together, and then reverse near $\sim$8110~days, moving inward.  In the $\xi = 100$ model, the overall structures are less coherent, but the ingoing pattern after $\sim$8110~days is still visible.  It is arguably still visible in the $\xi =1000$ model.  Interestingly, the high $\xi$ models do not appear to resemble the lamppost-like models that we see in the highly-smoothed models of the other AGNs. Using data from later epochs as well as the ones we use, \citet{vincentelli21} observe similar longer-term variability that is not consistent with the lamppost model, though they attribute this to be ``contamination'' by the diffuse continuum emission of the BLR.  


\subsection{NGC~4151}

In Figure \ref{fig:ngc4151_panel}, we show the temperature maps constructed using $N_t = 100$ and $\xi$ ranging from 1 to 1000.  In Figure \ref{fig:ngc4151_lc}, we show the observed lightcurves and the model lightcurves constructed using $\xi=10$.  Compared to the other AGNs, the main structures appear relatively stable across the full range of smoothing $\xi$.  The most prominent feature of the map, alternating hot and cold patches between $u=10$ and $u=100$, appears even with high degrees of smoothing ($\xi = 100$). 


\subsection{NGC~4593}

In Figure \ref{fig:ngc4593_panel}, we show the temperature maps constructed using $N_t = 100$ and $\xi$ ranging from 1 to 1000.  In Figure \ref{fig:ngc4593_lc}, we show the observed lightcurves and the model lightcurves constructed using $\xi=100$.  Even at low $\xi$, there is no distinct pattern in the temperature fluctuations, though it is worth noting that below $\xi = 1000$, there are radial fluctuations that are not possible in the lamppost model.  The data span only $\sim$30~d, which is a much shorter time range than the span of the data for the other AGNs where we observe more interesting patterns.


\subsection{Mrk~509}

In Figure \ref{fig:mrk509_panel}, we show the temperature maps constructed using $N_t = 100$ and $\xi$ ranging from 1 to 1000.  In Figure \ref{fig:mrk509_lc}, we show the observed lightcurves and the model lightcurves constructed using $\xi=10$.  With $\xi\leq10$, we can see the alternating bands of positive and negative fluctuations that shift over time.  It is not clear if these bands move outward or inward, as the bands seem to shift relatively quickly compared to the $\sim$50~d periods where they remain stationary.  There is some evidence for an ingoing pattern between 8000--8050~days, though this pattern is not apparent at higher levels of smoothing.  Even at $\xi=1000$, there are still radial fluctuations that do not match the lamppost model.

In the highly-smoothed $\xi =1000$ models of Mrk~509, as well as in NGC~4151 and NGC~4593, we can see that there is often a concentration of radial structure around the ``middle'' of the radial space (the exact radii depend on the AGNs' physical parameters, see Fig.~\ref{fig:kernels}).  This is likely due to the specifics of the data.  As discussed earlier, there are only \swift~data for these objects, whereas other AGN have data in other filters from ground-based observations. With only five filters, only a limited radial region of the disc is well-constrained.


\section{Discussion}\label{sec:discussion}

We analyse the lightcurves of NGC~5548 and six other AGNs using our model of axisymmetric temperature fluctuations and find evidence for temperature fluctuations that do not match the lamppost model.  Several of these AGNs, in particular NGC~5548, show evidence for slow ingoing and outgoing waves.  While the details of the fluctuation patterns differ between AGN and depend on the degree of smoothing, there are some commonalities.  The structures often appear as radial bands with alternating positive and negative temperature fluctuations.  These bands move slowly (i.e., $v \ll c$) inward or outward over time, though there are times where the bands will move relatively quickly while still maintaining the banded structure.  Interestingly, these bands frequently have similar logarithmic radial width of order 0.3~dex not just for a given AGN, but for the different AGN. 

\begin{figure*}
\includegraphics[width=\linewidth]{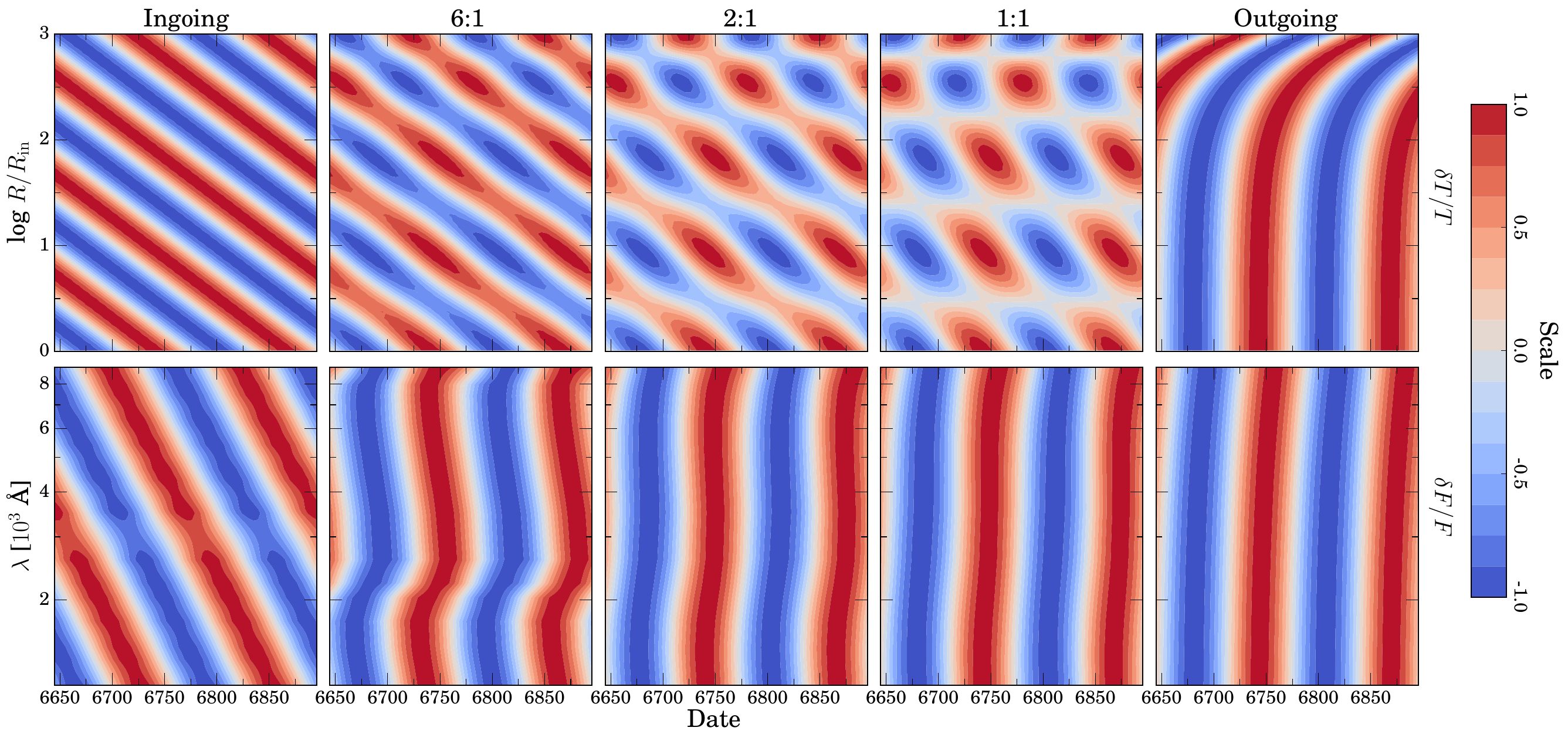}
\caption{Temperature ($\delta T/T$, top) and flux ($\delta F/F$, bottom) maps for various test scenarios.  The vertical axis of the $\delta F/F$ plots gives the wavelength of the ``observations''.  From left to right, we move from a pure slow ingoing wave to a pure fast outgoing wave.  In the 6:1, 2:1, and 1:1 columns, the amplitude of the ingoing wave is 6, 2, and 1 times the amplitude of the outgoing wave.  $\delta T/T$ is scaled as in previous figures.  $\delta F/F$ is scaled like the lightcurve figures with the minimum and maximum scaled to --1 and +1. }
\label{fig:smooth_panel}
\end{figure*}

As we increase the smoothing, the signals increasingly look like rapid ($v \sim c$) outgoing waves -- the lamppost model.  This is to be expected since the lightcurves do resemble time-shifted versions of each other with the longer wavelengths varying later than the shorter wavelengths by typical light travel times.   However, these heavily smoothed models are also poorer fits to the data and still show structure beyond a pure lamppost model. 

In these inversion problems, structures can be driven by noise in the data, and the goodness of fit statistic depends on the noise. However, we have tried to ensure that the noise estimates are realistic given the properties of the lightcurves (i.e., inflating the errors when necessary according to the triplet test in Sec.~\ref{sec:agndata}).  We believe that much of these structures are real.  All of the systems show extra structure, and much of the structure exists over broad ranges of the smoothing parameter.  Moreover, it would be physically unrealistic for additional temperature fluctuations to be totally absent.

In fact, most models of the origins of AGN variability invoke thermal fluctuations near the radii producing the observed photons (e.g., \citealt{kelly09,dexter11}) which are probed in our models, and certainly in simulations, the discs show significant  variability at these radii (e.g., \citealt{jiang19,jiang20}).  As noted earlier, the time-scales of the variability are also typical of thermal time-scales at these radii and much longer than any characteristic time-scale for the regions near the inner edge of the disc. So how can the temperature variability be dominated by local thermal fluctuations yet produce lightcurves that resemble the lamppost model?

The answer is that the radial smoothing effects of the blackbody kernel (Eqs.~\ref{eq:flux} and \ref{eq:dfdt}) can exponentially suppress the observed contributions from slow moving waves compared to fast waves like a lamppost signal.  Nothing is less washed-out than a lamppost-like signal because it is moving at the speed of light. As a simple model problem, we consider a wave 
\begin{equation}
F(t,r) = A \sin{(\omega t \pm kr)} = A \sin{\bigg(\omega t \pm \frac{2\pi}{P v} r \bigg)}
\end{equation}
where $v = k\omega$ is the phase velocity and $P = 2\pi/\omega$ is the period.  If we smooth the wave in $r$ with a Gaussian of the form 
\begin{equation}
S(r) = \exp{\bigg(-\frac{1}{2}\bigg[\frac{r-r_0}{\varepsilon r_0}\bigg]^2 \bigg) } 
\end{equation}
where $r_0$ is the center of the kernel and $\varepsilon r_0$ is the width of the kernel in terms of the fraction $\varepsilon$ of $r_0$ ($\varepsilon$ is $\sim$0.6 for Eq.~\ref{eq:dfdt}), the observed wave amplitude is
\begin{equation}
A \exp{\bigg( -\frac{1}{2} \bigg[\frac{2\pi\varepsilon r_0}{Pv}\bigg]^2 \bigg) } ~.
\end{equation}
where
\begin{equation} \label{eq:smoothing}
\frac{2\pi\varepsilon r_0}{Pv} = 0.072 \bigg(\frac{\varepsilon r_0}{600~R_g}\bigg) \bigg(\frac{30~\rm{d}}{P}\bigg) \bigg(\frac{M_{\rm BH}}{10^7~\rm{M_\odot}}\bigg)  \bigg(\frac{c}{v}\bigg) ~.
\end{equation} 
A wave moving with $v\ll c$ is going to be exponentially more damped than a wave moving at $v\sim c$. Given a characteristic variability time-scale of order a month and the $M_{\rm BH}$ of most of these targets, we would expect that a lamppost signal/lag is more easily seen in the lightcurves over the slow moving waves that we see in our models.  Note that the degree of damping depends only on the propagation speed of the perturbation and not on the direction of propagation (ingoing or outgoing).

We can also evaluate this scenario using test problems. In Figure \ref{fig:smooth_panel}, we synthesize five test scenarios (two of which are the same as those in Section~\ref{sec:testing}, and one of which is considered in Appendix~\ref{sec:appendix}) -- starting from a pure slow ingoing wave and then adding a fast (0.1$c$) outgoing wave of increasing relative strength from 6:1 to 2:1 to 1:1 and then ending with a pure fast outgoing wave.  Below these temperature fluctuation maps, we also show the directly-integrated differential fluxes $\delta F/F$ of these test scenarios in the form of color maps.  These fluxes are calculated as in Section~\ref{sec:testing} but using a uniform grid in time.  We do not include errors or noise in the integration, and $\delta F/F$ is scaled to the range --1 to +1.  

The flux maps for the mixed models closely resemble those of the pure fast outgoing model even when the slow ingoing wave is 6 times larger in amplitude.  This visually illustrates the damping of the flux contribution from slow waves compared to fast waves.  Note that the problem is even more severe for non-axisymmetric temperature fluctuations because they are smoothed in azimuth as well as radius.  This effect provides a natural explanation of how disc-driven temperature fluctuations and slow-moving waves can dominate the temperature variability while the lightcurves are dominated by a lamppost signal.  

In fact, one could use slow higher-amplitude ingoing waves to drive a lamppost signal.  As the ingoing temperature perturbations approach the inner radii of the disc, they could create a ``disc-driven lamppost'' with the slower time-scales of larger disc radii.  This idea is not dissimilar to the ideas in \citet{lyubarskii97} (see also \citealt{kotov01,uttley01}), where viscosity fluctuations in the outer disc drive changes in the accretion rates/luminosity of the inner disc.  The disc-driven lamppost then irradiates the disc to produce a fast lower-amplitude outgoing signal.  Because of the smoothing properties of the disc, the lightcurves are dominated by this secondary signal even though it is a lower amplitude temperature perturbation than the driving ingoing wave.  

Moreover, longer-term variability that cannot be explained by the lamppost model appears to become more prominent in lightcurves with longer temporal baselines, as in the case of Fairall~9 and Mrk~110 (\citealt{hernandez20} and \citealt{vincentelli21}, respectively).  This is somewhat expected, in that longer time-scales are likely associated with larger physical scales, making the smoothing by the blackbody kernel less effective at damping the signal.  Because the data used here all comes from short duration RM programs, they provide little information on the wavelength dependence of longer-time-scale variability.

A good next target for our studies is Mrk~817 using data similar to that of NGC~5548 being obtained as part of AGN STORM 2 \citep{kara21}.  While it is not feasible to obtain lightcurves further into the UV than the Lyman limit, near-IR data would significantly extend the coverage to larger radii in future programs.  Finally, our models could be used on a massive scale to analyse data from the upcoming Vera C. Rubin Observatory (Rubin, \citealt{ivezic19}). While the typical cadence of Rubin observations is not ideal for disc RM studies of objects similar to the ones considered here, Rubin can likely do good RM studies of higher mass and redshift systems where the increased size and cosmologically slowed variability compensate for the slower observing cadence.  The wavelength range spanned by Rubin explores a similar radial range to that of the \swift-only systems considered here in the sense that the ratio of the minimum and maximum filter wavelengths sets the ratio of the inner and outer radii that can be explored ($u_{\rm max}/u_{\rm min} \propto (\lambda_{\rm max}/\lambda_{\rm min})^{4/3}$ for a thin disc).  Perhaps more importantly, Rubin will operate for far longer than the RM campaigns considered here and will provide the first well-characterized multiwavelength lightcurves to explore what happens on long time-scales and whether variability on long time-scales looks increasingly less like a lamppost model.  The potential is great since Rubin will provide potentially useful lightcurves for hundreds of thousands to millions of AGNs.

\section{Acknowledgements}

We thank S.~Mathur, P.~Martini, and Z.~Yu for providing essential feedback on our work.  We thank F.~Vincentelli for providing us with the reduced data for Mrk~110.  JMMN and CSK are supported by NSF grants AST-1814440 and AST-1908570.

\section{Online supplementary material}

We include animated versions of the temperature fluctuation maps as a function of $\xi$ for the various AGNs.  These versions show in finer detail how the fluctuation patterns and $\chi^2$ change as $\xi$ is increased (and decreased).  Because they are animated, they are included only as online supplementary material.  

\section{Data availability statement}

All data used in this paper are publically available.

\appendix
\section{Additional test cases}\label{sec:appendix}

\begin{figure*}
\includegraphics[width=\linewidth]{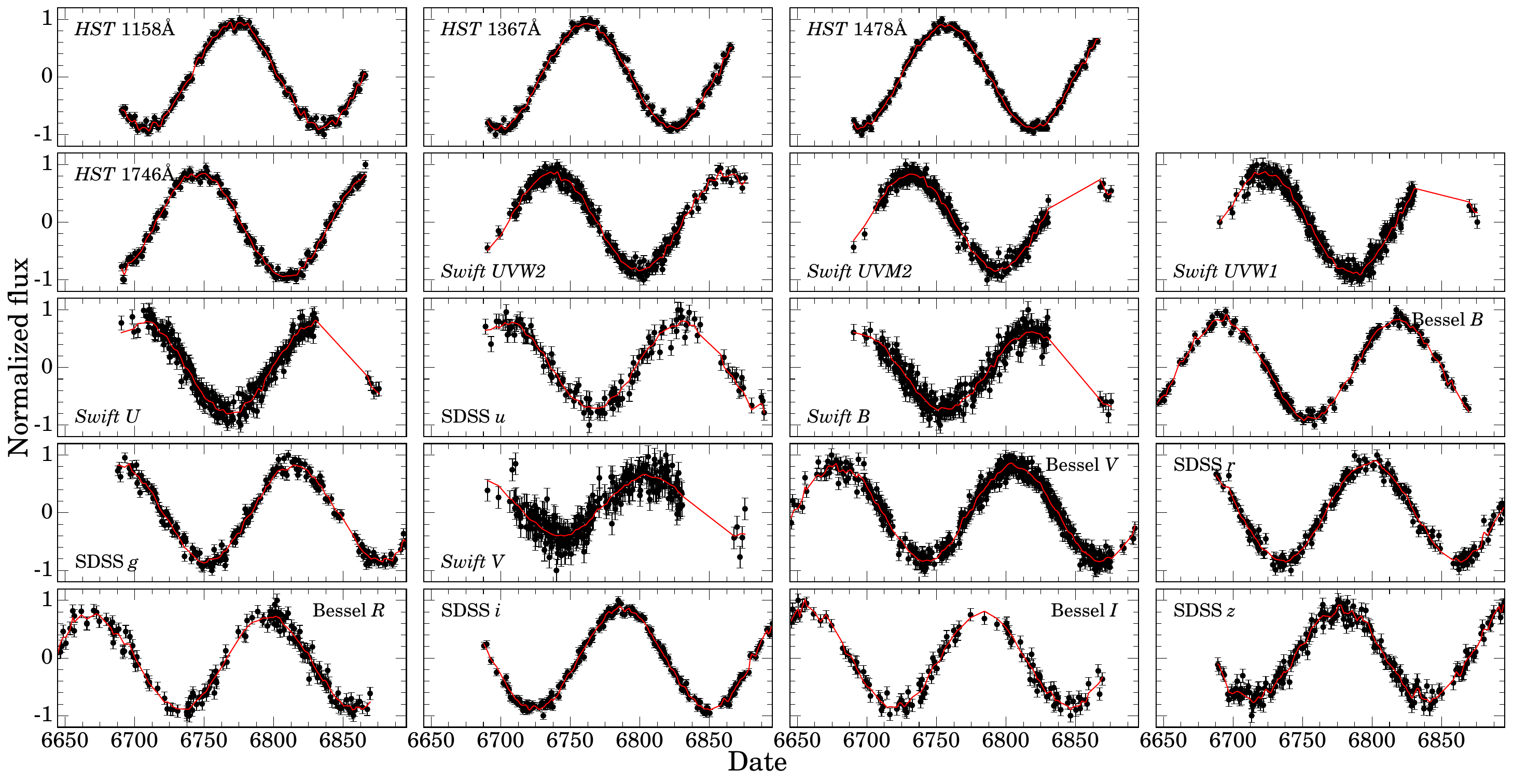}
\caption{Input (black) and model (red) lightcurves for the \textit{ingo} (slow ingoing) test pattern with $N_t = 100$ and $\xi=10$.  See Fig.~\ref{fig:ingo_panel} for the input and reconstructed temperature fluctuation patterns.}
\label{fig:ingo_lc}
\end{figure*}

In Section~\ref{sec:testing}, we tested our model with two temperature fluctuation patterns -- \textit{ingo} and \textit{outgo}.  In this Appendix, we show the synthetic and reconstructed lightcurves for the \textit{ingo} pattern in Figure~\ref{fig:ingo_lc}, which was omitted from Section~\ref{sec:testing}.

\begin{figure*}
\includegraphics[width=\linewidth]{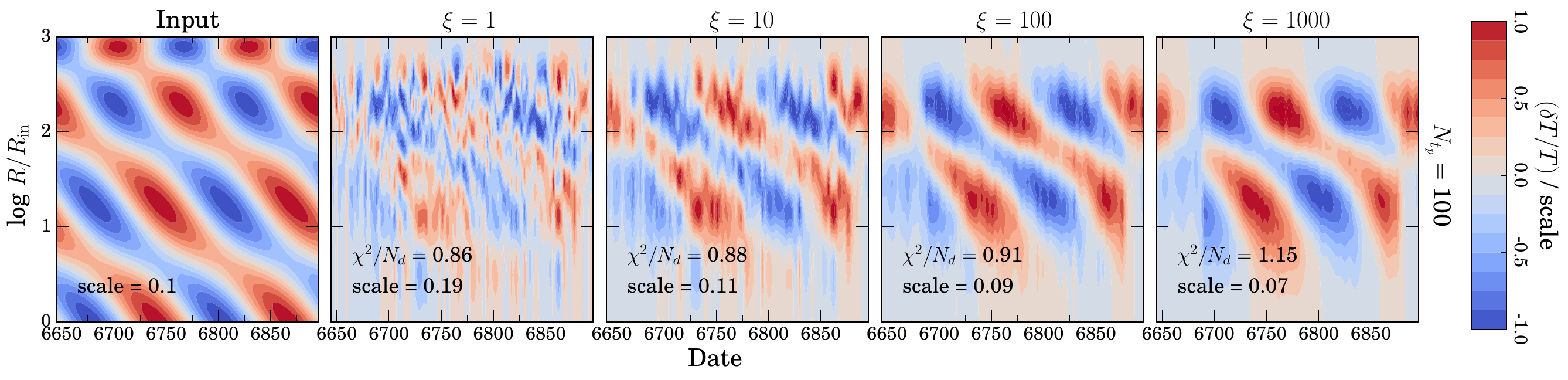}
\caption{Input and reconstructed temperature maps for the \textit{in-and-out} test pattern with $N_t = 100$ and $\xi = 1,10,100,1000$.}
\label{fig:innout_panel}
\end{figure*}

\begin{figure*}
\includegraphics[width=\linewidth]{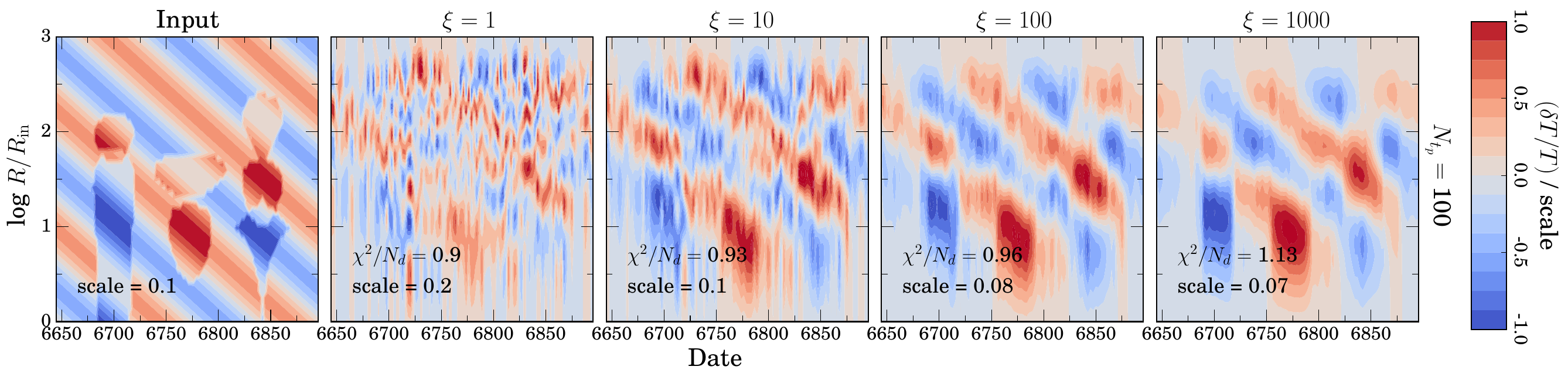}
\caption{Input and reconstructed temperature maps for the \textit{bumpy} test pattern with $N_t = 100$ and $\xi = 1,10,100,1000$.}
\label{fig:bumps_panel}
\end{figure*}

\begin{figure*}
\includegraphics[width=\linewidth]{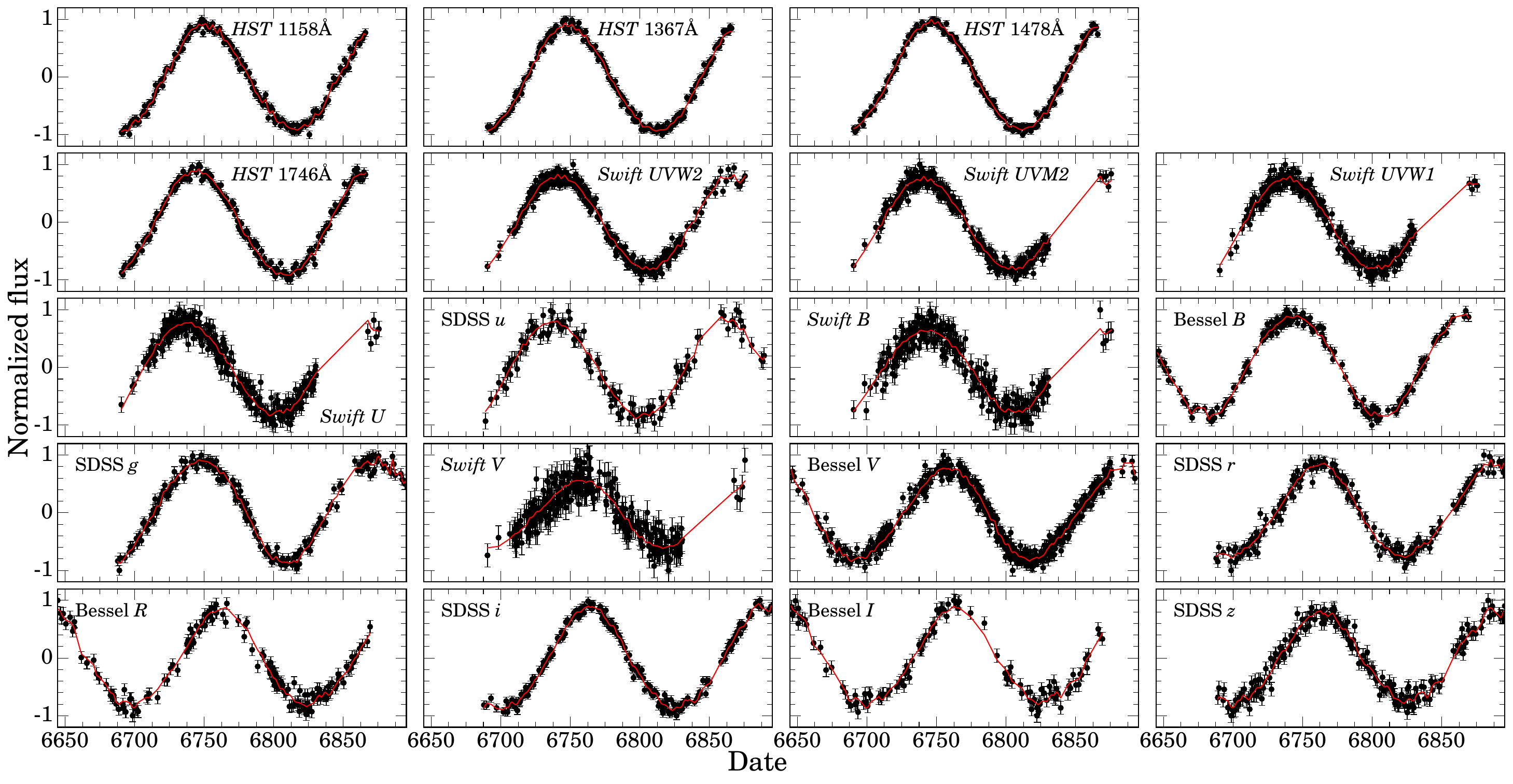}
\caption{Input (black) and model (red) lightcurves for the \textit{in-and-out} test pattern with $N_t = 100$ and $\xi=10$.}
\label{fig:innout_lc}
\end{figure*}

\begin{figure*}
\includegraphics[width=\linewidth]{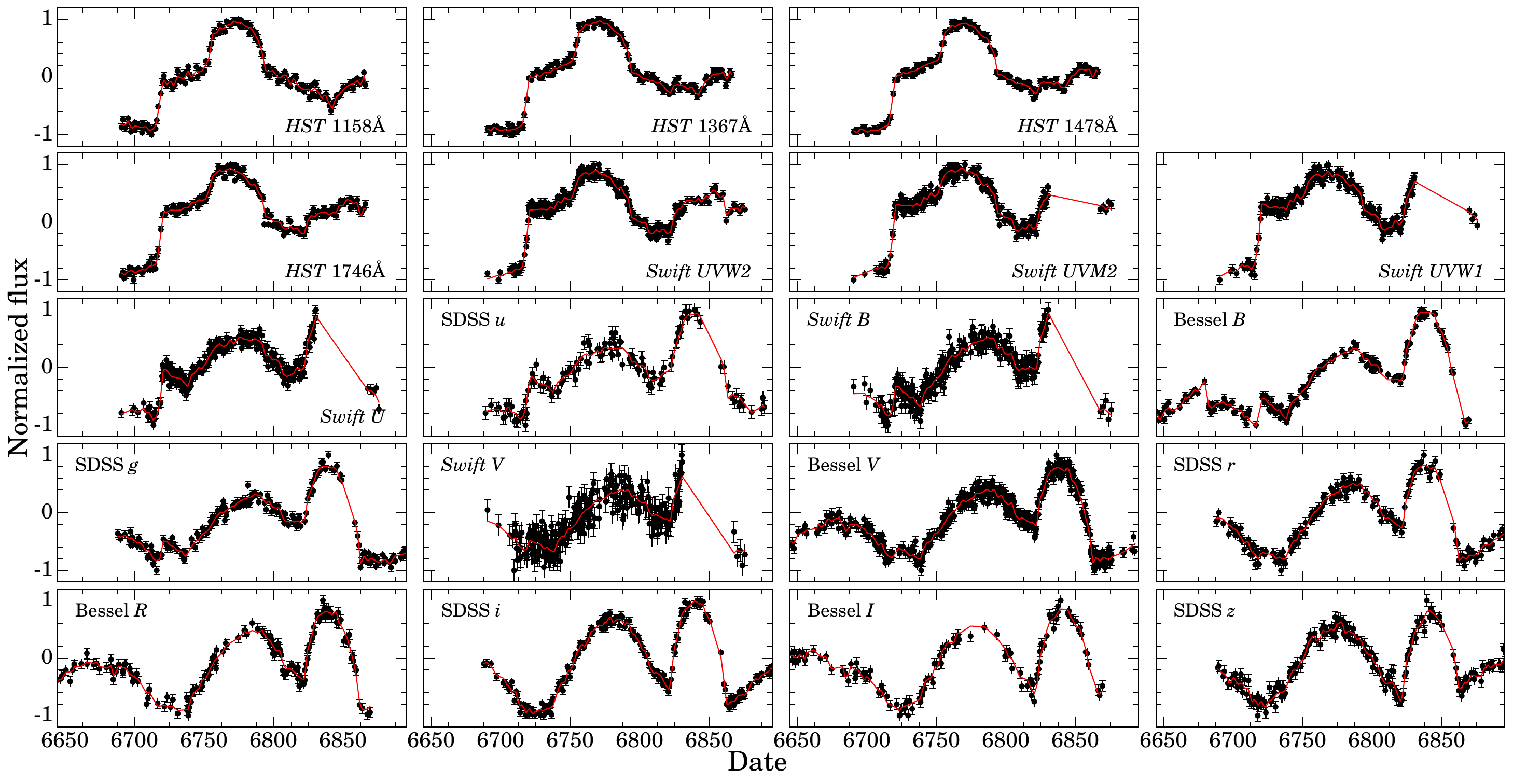}
\caption{Input (black) and model (red) lightcurves for the \textit{bumpy} test pattern with $N_t = 100$ and $\xi=10$.}
\label{fig:bumps_lc}
\end{figure*}

We also test an additional two patterns:  \textit{in-and-out},  where we combine the \textit{ingo} pattern and the \textit{outgo} pattern such that the amplitude of the \textit{ingo} is twice that of the \textit{outgo}, and \textit{bumps}, where temperature fluctuations are placed at arbitrary points in radius and time on top of the \textit{ingo} pattern.  The input and reconstructed temperature fluctuation patterns for \textit{in-and-out} and \textit{bumps} are shown as Figures~\ref{fig:innout_panel} and \ref{fig:bumps_panel}, respectively, and the synthetic and reconstructed lightcurves are shown as Figures~\ref{fig:innout_lc} and \ref{fig:bumps_lc}, respectively.  

The purpose of these additional two exercises are to show how well our model reconstructs test fluctuations in spite of potentially peculiar fluctuations -- like the \textit{bumps} model.  Furthermore, the \textit{in-and-out} pattern is the same as one of the combination fast outgoing and slow ingoing patterns considered in Section~\ref{sec:discussion}.

\bibliographystyle{mnras}
\bibliography{bibliography}

\begin{thebibliography}{}
\makeatletter
\relax
\def\mn@urlcharsother{\let\do\@makeother \do\$\do\&\do\#\do\^\do\_\do\%\do\~}
\def\mn@doi{\begingroup\mn@urlcharsother \@ifnextchar [ {\mn@doi@}
  {\mn@doi@[]}}
\def\mn@doi@[#1]#2{\def\@tempa{#1}\ifx\@tempa\@empty \href
  {http://dx.doi.org/#2} {doi:#2}\else \href {http://dx.doi.org/#2} {#1}\fi
  \endgroup}
\def\mn@eprint#1#2{\mn@eprint@#1:#2::\@nil}
\def\mn@eprint@arXiv#1{\href {http://arxiv.org/abs/#1} {{\tt arXiv:#1}}}
\def\mn@eprint@dblp#1{\href {http://dblp.uni-trier.de/rec/bibtex/#1.xml}
  {dblp:#1}}
\def\mn@eprint@#1:#2:#3:#4\@nil{\def\@tempa {#1}\def\@tempb {#2}\def\@tempc
  {#3}\ifx \@tempc \@empty \let \@tempc \@tempb \let \@tempb \@tempa \fi \ifx
  \@tempb \@empty \def\@tempb {arXiv}\fi \@ifundefined
  {mn@eprint@\@tempb}{\@tempb:\@tempc}{\expandafter \expandafter \csname
  mn@eprint@\@tempb\endcsname \expandafter{\@tempc}}}

\bibitem[\protect\citeauthoryear{{Berkley}, {Kazanas}  \& {Ozik}}{{Berkley}
  et~al.}{2000}]{berkley00}
{Berkley} A.~J.,  {Kazanas} D.,   {Ozik} J.,  2000, \mn@doi [\apj]
  {10.1086/308880}, \href
  {https://ui.adsabs.harvard.edu/abs/2000ApJ...535..712B} {535, 712}

\bibitem[\protect\citeauthoryear{{Blandford} \& {McKee}}{{Blandford} \&
  {McKee}}{1982}]{blandford82}
{Blandford} R.~D.,  {McKee} C.~F.,  1982, \mn@doi [\apj] {10.1086/159843},
  \href {https://ui.adsabs.harvard.edu/abs/1982ApJ...255..419B} {255, 419}

\bibitem[\protect\citeauthoryear{{Burke} et~al.,}{{Burke}
  et~al.}{2021}]{burke21}
{Burke} C.~J.,  et~al., 2021, \mn@doi [Science] {10.1126/science.abg9933},
  \href {https://ui.adsabs.harvard.edu/abs/2021Sci...373..789B} {373, 789}

\bibitem[\protect\citeauthoryear{{Cackett}, {Horne}  \& {Winkler}}{{Cackett}
  et~al.}{2007}]{cackett07}
{Cackett} E.~M.,  {Horne} K.,   {Winkler} H.,  2007, \mn@doi [\mnras]
  {10.1111/j.1365-2966.2007.12098.x}, \href
  {https://ui.adsabs.harvard.edu/abs/2007MNRAS.380..669C} {380, 669}

\bibitem[\protect\citeauthoryear{{Cackett} et~al.,}{{Cackett}
  et~al.}{2020}]{cackett20}
{Cackett} E.~M.,  et~al., 2020, \mn@doi [\apj] {10.3847/1538-4357/ab91b5},
  \href {https://ui.adsabs.harvard.edu/abs/2020ApJ...896....1C} {896, 1}

\bibitem[\protect\citeauthoryear{{Cristiani}, {Trentini}, {La Franca}  \&
  {Andreani}}{{Cristiani} et~al.}{1997}]{cristiani97}
{Cristiani} S.,  {Trentini} S.,  {La Franca} F.,   {Andreani} P.,  1997, \aap,
  \href {https://ui.adsabs.harvard.edu/abs/1997A&A...321..123C} {321, 123}

\bibitem[\protect\citeauthoryear{{De Rosa} et~al.,}{{De Rosa}
  et~al.}{2015}]{derosa15}
{De Rosa} G.,  et~al., 2015, \mn@doi [\apj] {10.1088/0004-637X/806/1/128},
  \href {https://ui.adsabs.harvard.edu/abs/2015ApJ...806..128D} {806, 128}

\bibitem[\protect\citeauthoryear{{Dexter} \& {Agol}}{{Dexter} \&
  {Agol}}{2011}]{dexter11}
{Dexter} J.,  {Agol} E.,  2011, \mn@doi [\apjl] {10.1088/2041-8205/727/1/L24},
  \href {https://ui.adsabs.harvard.edu/abs/2011ApJ...727L..24D} {727, L24}

\bibitem[\protect\citeauthoryear{{Dexter} et~al.,}{{Dexter}
  et~al.}{2019}]{dexter19}
{Dexter} J.,  et~al., 2019, \mn@doi [\apj] {10.3847/1538-4357/ab4354}, \href
  {https://ui.adsabs.harvard.edu/abs/2019ApJ...885...44D} {885, 44}

\bibitem[\protect\citeauthoryear{{Edelson} et~al.,}{{Edelson}
  et~al.}{2015}]{edelson15}
{Edelson} R.,  et~al., 2015, \mn@doi [\apj] {10.1088/0004-637X/806/1/129},
  \href {https://ui.adsabs.harvard.edu/abs/2015ApJ...806..129E} {806, 129}

\bibitem[\protect\citeauthoryear{{Edelson} et~al.,}{{Edelson}
  et~al.}{2017}]{edelson17}
{Edelson} R.,  et~al., 2017, \mn@doi [\apj] {10.3847/1538-4357/aa6890}, \href
  {https://ui.adsabs.harvard.edu/abs/2017ApJ...840...41E} {840, 41}

\bibitem[\protect\citeauthoryear{{Edelson} et~al.,}{{Edelson}
  et~al.}{2019}]{edelson19}
{Edelson} R.,  et~al., 2019, \mn@doi [\apj] {10.3847/1538-4357/aaf3b4}, \href
  {https://ui.adsabs.harvard.edu/abs/2019ApJ...870..123E} {870, 123}

\bibitem[\protect\citeauthoryear{{Fausnaugh} et~al.,}{{Fausnaugh}
  et~al.}{2016}]{fausnaugh16}
{Fausnaugh} M.~M.,  et~al., 2016, \mn@doi [\apj] {10.3847/0004-637X/821/1/56},
  \href {https://ui.adsabs.harvard.edu/abs/2016ApJ...821...56F} {821, 56}

\bibitem[\protect\citeauthoryear{{Frank}, {King}  \& {Raine}}{{Frank}
  et~al.}{2002}]{frank02}
{Frank} J.,  {King} A.,   {Raine} D.~J.,  2002, {Accretion Power in
  Astrophysics: Third Edition}.
Cambridge University Press

\bibitem[\protect\citeauthoryear{{Geha} et~al.,}{{Geha} et~al.}{2003}]{geha03}
{Geha} M.,  et~al., 2003, \mn@doi [\aj] {10.1086/344947}, \href
  {https://ui.adsabs.harvard.edu/abs/2003AJ....125....1G} {125, 1}

\bibitem[\protect\citeauthoryear{{Gehrels} et~al.,}{{Gehrels}
  et~al.}{2004}]{gehrels04}
{Gehrels} N.,  et~al., 2004, \mn@doi [\apj] {10.1086/422091}, \href
  {http://adsabs.harvard.edu/abs/2004ApJ...611.1005G} {611, 1005}

\bibitem[\protect\citeauthoryear{{Giveon}, {Maoz}, {Kaspi}, {Netzer}  \&
  {Smith}}{{Giveon} et~al.}{1999}]{giveon99}
{Giveon} U.,  {Maoz} D.,  {Kaspi} S.,  {Netzer} H.,   {Smith} P.~S.,  1999,
  \mn@doi [\mnras] {10.1046/j.1365-8711.1999.02556.x}, \href
  {https://ui.adsabs.harvard.edu/abs/1999MNRAS.306..637G} {306, 637}

\bibitem[\protect\citeauthoryear{{Hern{\'a}ndez Santisteban}
  et~al.,}{{Hern{\'a}ndez Santisteban} et~al.}{2020}]{hernandez20}
{Hern{\'a}ndez Santisteban} J.~V.,  et~al., 2020, \mn@doi [\mnras]
  {10.1093/mnras/staa2365}, \href
  {https://ui.adsabs.harvard.edu/abs/2020MNRAS.498.5399H} {498, 5399}

\bibitem[\protect\citeauthoryear{{Ivezi{\'c}} et~al.,}{{Ivezi{\'c}}
  et~al.}{2019}]{ivezic19}
{Ivezi{\'c}} {\v{Z}}.,  et~al., 2019, \mn@doi [\apj]
  {10.3847/1538-4357/ab042c}, \href
  {https://ui.adsabs.harvard.edu/abs/2019ApJ...873..111I} {873, 111}

\bibitem[\protect\citeauthoryear{{Jiang} \& {Blaes}}{{Jiang} \&
  {Blaes}}{2020}]{jiang20}
{Jiang} Y.-F.,  {Blaes} O.,  2020, \mn@doi [\apj] {10.3847/1538-4357/aba4b7},
  \href {https://ui.adsabs.harvard.edu/abs/2020ApJ...900...25J} {900, 25}

\bibitem[\protect\citeauthoryear{{Jiang}, {Blaes}, {Stone}  \& {Davis}}{{Jiang}
  et~al.}{2019}]{jiang19}
{Jiang} Y.-F.,  {Blaes} O.,  {Stone} J.~M.,   {Davis} S.~W.,  2019, \mn@doi
  [\apj] {10.3847/1538-4357/ab4a00}, \href
  {https://ui.adsabs.harvard.edu/abs/2019ApJ...885..144J} {885, 144}

\bibitem[\protect\citeauthoryear{{Kara} et~al.,}{{Kara} et~al.}{2021}]{kara21}
{Kara} E.,  et~al., 2021, \mn@doi [\apj] {10.3847/1538-4357/ac2159}, \href
  {https://ui.adsabs.harvard.edu/abs/2021ApJ...922..151K} {922, 151}

\bibitem[\protect\citeauthoryear{{Kazanas} \& {Nayakshin}}{{Kazanas} \&
  {Nayakshin}}{2001}]{kazanas01}
{Kazanas} D.,  {Nayakshin} S.,  2001, \mn@doi [\apj] {10.1086/319786}, \href
  {https://ui.adsabs.harvard.edu/abs/2001ApJ...550..655K} {550, 655}

\bibitem[\protect\citeauthoryear{{Kelly}, {Bechtold}  \&
  {Siemiginowska}}{{Kelly} et~al.}{2009}]{kelly09}
{Kelly} B.~C.,  {Bechtold} J.,   {Siemiginowska} A.,  2009, \mn@doi [\apj]
  {10.1088/0004-637X/698/1/895}, \href
  {https://ui.adsabs.harvard.edu/abs/2009ApJ...698..895K} {698, 895}

\bibitem[\protect\citeauthoryear{{Kelly}, {Sobolewska}  \&
  {Siemiginowska}}{{Kelly} et~al.}{2011}]{kelly11}
{Kelly} B.~C.,  {Sobolewska} M.,   {Siemiginowska} A.,  2011, \mn@doi [\apj]
  {10.1088/0004-637X/730/1/52}, \href
  {https://ui.adsabs.harvard.edu/abs/2011ApJ...730...52K} {730, 52}

\bibitem[\protect\citeauthoryear{{Korista} \& {Goad}}{{Korista} \&
  {Goad}}{2001}]{korista01}
{Korista} K.~T.,  {Goad} M.~R.,  2001, \mn@doi [\apj] {10.1086/320964}, \href
  {https://ui.adsabs.harvard.edu/abs/2001ApJ...553..695K} {553, 695}

\bibitem[\protect\citeauthoryear{{Kotov}, {Churazov}  \& {Gilfanov}}{{Kotov}
  et~al.}{2001}]{kotov01}
{Kotov} O.,  {Churazov} E.,   {Gilfanov} M.,  2001, \mn@doi [\mnras]
  {10.1046/j.1365-8711.2001.04769.x}, \href
  {https://ui.adsabs.harvard.edu/abs/2001MNRAS.327..799K} {327, 799}

\bibitem[\protect\citeauthoryear{{Koz{\l}owski} et~al.,}{{Koz{\l}owski}
  et~al.}{2010}]{kozlowski10}
{Koz{\l}owski} S.,  et~al., 2010, \mn@doi [\apj] {10.1088/0004-637X/708/2/927},
  \href {https://ui.adsabs.harvard.edu/abs/2010ApJ...708..927K} {708, 927}

\bibitem[\protect\citeauthoryear{{Lawther}, {Goad}, {Korista}, {Ulrich}  \&
  {Vestergaard}}{{Lawther} et~al.}{2018}]{lawther18}
{Lawther} D.,  {Goad} M.~R.,  {Korista} K.~T.,  {Ulrich} O.,   {Vestergaard}
  M.,  2018, \mn@doi [\mnras] {10.1093/mnras/sty2242}, \href
  {https://ui.adsabs.harvard.edu/abs/2018MNRAS.481..533L} {481, 533}

\bibitem[\protect\citeauthoryear{{Li} et~al.,}{{Li} et~al.}{2018}]{li18}
{Li} Y.-R.,  et~al., 2018, \mn@doi [\apj] {10.3847/1538-4357/aaee6b}, \href
  {https://ui.adsabs.harvard.edu/abs/2018ApJ...869..137L} {869, 137}

\bibitem[\protect\citeauthoryear{{Lyubarskii}}{{Lyubarskii}}{1997}]{lyubarskii97}
{Lyubarskii} Y.~E.,  1997, \mn@doi [\mnras] {10.1093/mnras/292.3.679}, \href
  {https://ui.adsabs.harvard.edu/abs/1997MNRAS.292..679L} {292, 679}

\bibitem[\protect\citeauthoryear{{MacLeod} et~al.,}{{MacLeod}
  et~al.}{2010}]{macleod10}
{MacLeod} C.~L.,  et~al., 2010, \mn@doi [\apj] {10.1088/0004-637X/721/2/1014},
  \href {https://ui.adsabs.harvard.edu/abs/2010ApJ...721.1014M} {721, 1014}

\bibitem[\protect\citeauthoryear{{MacLeod} et~al.,}{{MacLeod}
  et~al.}{2012}]{macleod12}
{MacLeod} C.~L.,  et~al., 2012, \mn@doi [\apj] {10.1088/0004-637X/753/2/106},
  \href {https://ui.adsabs.harvard.edu/abs/2012ApJ...753..106M} {753, 106}

\bibitem[\protect\citeauthoryear{{Mushotzky}, {Edelson}, {Baumgartner}  \&
  {Gandhi}}{{Mushotzky} et~al.}{2011}]{mushotzky11}
{Mushotzky} R.~F.,  {Edelson} R.,  {Baumgartner} W.,   {Gandhi} P.,  2011,
  \mn@doi [\apjl] {10.1088/2041-8205/743/1/L12}, \href
  {https://ui.adsabs.harvard.edu/abs/2011ApJ...743L..12M} {743, L12}

\bibitem[\protect\citeauthoryear{{Nayakshin}, {Kazanas}  \&
  {Kallman}}{{Nayakshin} et~al.}{2000}]{nayakshin00}
{Nayakshin} S.,  {Kazanas} D.,   {Kallman} T.~R.,  2000, \mn@doi [\apj]
  {10.1086/309054}, \href
  {https://ui.adsabs.harvard.edu/abs/2000ApJ...537..833N} {537, 833}

\bibitem[\protect\citeauthoryear{{Oknyanskij}}{{Oknyanskij}}{1978}]{oknyanskij78}
{Oknyanskij} V.~L.,  1978, Peremennye Zvezdy, \href
  {https://ui.adsabs.harvard.edu/abs/1978PZ.....21...71O} {21, 71}

\bibitem[\protect\citeauthoryear{{Perola} et~al.,}{{Perola}
  et~al.}{1982}]{perola82}
{Perola} G.~C.,  et~al., 1982, \mn@doi [\mnras] {10.1093/mnras/200.2.293},
  \href {https://ui.adsabs.harvard.edu/abs/1982MNRAS.200..293P} {200, 293}

\bibitem[\protect\citeauthoryear{{Peterson}}{{Peterson}}{1993}]{peterson93}
{Peterson} B.~M.,  1993, \mn@doi [\pasp] {10.1086/133140}, \href
  {https://ui.adsabs.harvard.edu/abs/1993PASP..105..247P} {105, 247}

\bibitem[\protect\citeauthoryear{{Roming} et~al.,}{{Roming}
  et~al.}{2005}]{roming05}
{Roming} P.~W.~A.,  et~al., 2005, \mn@doi [SSR] {10.1007/s11214-005-5095-4},
  \href {http://adsabs.harvard.edu/abs/2005SSRv..120...95R} {120, 95}

\bibitem[\protect\citeauthoryear{{Rumbaugh} et~al.,}{{Rumbaugh}
  et~al.}{2018}]{rumbaugh18}
{Rumbaugh} N.,  et~al., 2018, \mn@doi [\apj] {10.3847/1538-4357/aaa9b6}, \href
  {https://ui.adsabs.harvard.edu/abs/2018ApJ...854..160R} {854, 160}

\bibitem[\protect\citeauthoryear{{Sergeev}, {Doroshenko}, {Golubinskiy},
  {Merkulova}  \& {Sergeeva}}{{Sergeev} et~al.}{2005}]{sergeev05}
{Sergeev} S.~G.,  {Doroshenko} V.~T.,  {Golubinskiy} Y.~V.,  {Merkulova} N.~I.,
    {Sergeeva} E.~A.,  2005, \mn@doi [\apj] {10.1086/427820}, \href
  {https://ui.adsabs.harvard.edu/abs/2005ApJ...622..129S} {622, 129}

\bibitem[\protect\citeauthoryear{{Shakura} \& {Sunyaev}}{{Shakura} \&
  {Sunyaev}}{1973}]{shakura73}
{Shakura} N.~I.,  {Sunyaev} R.~A.,  1973, \aap, \href
  {https://ui.adsabs.harvard.edu/abs/1973A&A....24..337S} {500, 33}

\bibitem[\protect\citeauthoryear{{Shappee} et~al.,}{{Shappee}
  et~al.}{2014}]{shappee14}
{Shappee} B.~J.,  et~al., 2014, \mn@doi [\apj] {10.1088/0004-637X/788/1/48},
  \href {https://ui.adsabs.harvard.edu/abs/2014ApJ...788...48S} {788, 48}

\bibitem[\protect\citeauthoryear{{Starkey} et~al.,}{{Starkey}
  et~al.}{2017}]{starkey17}
{Starkey} D.,  et~al., 2017, \mn@doi [\apj] {10.3847/1538-4357/835/1/65}, \href
  {https://ui.adsabs.harvard.edu/abs/2017ApJ...835...65S} {835, 65}

\bibitem[\protect\citeauthoryear{{Ulrich}, {Maraschi}  \& {Urry}}{{Ulrich}
  et~al.}{1997}]{ulrich97}
{Ulrich} M.-H.,  {Maraschi} L.,   {Urry} C.~M.,  1997, \mn@doi [\araa]
  {10.1146/annurev.astro.35.1.445}, \href
  {https://ui.adsabs.harvard.edu/abs/1997ARA&A..35..445U} {35, 445}

\bibitem[\protect\citeauthoryear{{Uttley} \& {McHardy}}{{Uttley} \&
  {McHardy}}{2001}]{uttley01}
{Uttley} P.,  {McHardy} I.~M.,  2001, \mn@doi [\mnras]
  {10.1046/j.1365-8711.2001.04496.x}, \href
  {https://ui.adsabs.harvard.edu/abs/2001MNRAS.323L..26U} {323, L26}

\bibitem[\protect\citeauthoryear{{Vasudevan} \& {Fabian}}{{Vasudevan} \&
  {Fabian}}{2009}]{vasudevan09}
{Vasudevan} R.~V.,  {Fabian} A.~C.,  2009, \mn@doi [\mnras]
  {10.1111/j.1365-2966.2008.14108.x}, \href
  {https://ui.adsabs.harvard.edu/abs/2009MNRAS.392.1124V} {392, 1124}

\bibitem[\protect\citeauthoryear{{Vincentelli} et~al.,}{{Vincentelli}
  et~al.}{2021}]{vincentelli21}
{Vincentelli} F.~M.,  et~al., 2021, \mn@doi [\mnras] {10.1093/mnras/stab1033},
  \href {https://ui.adsabs.harvard.edu/abs/2021MNRAS.504.4337V} {504, 4337}

\bibitem[\protect\citeauthoryear{{Wright}}{{Wright}}{2006}]{wright06}
{Wright} E.~L.,  2006, \mn@doi [\pasp] {10.1086/510102}, \href
  {http://adsabs.harvard.edu/abs/2006PASP..118.1711W} {118, 1711}

\bibitem[\protect\citeauthoryear{{Zu}, {Kochanek}, {Koz{\l}owski}  \&
  {Udalski}}{{Zu} et~al.}{2013}]{zu13}
{Zu} Y.,  {Kochanek} C.~S.,  {Koz{\l}owski} S.,   {Udalski} A.,  2013, \mn@doi
  [\apj] {10.1088/0004-637X/765/2/106}, \href
  {https://ui.adsabs.harvard.edu/abs/2013ApJ...765..106Z} {765, 106}

\makeatother
\end{thebibliography}


\label{lastpage}
\bsp	
\end{document}